\newcommand{\beq}{\begin{equation}}
\newcommand{\eeq}{\end{equation}}
\newcommand{\dd}{{\rm d}}
\newcommand{\fig}[1]{Fig.\,\ref{#1}}
\newcommand{\eqn}[1]{Eq.\,(\ref{#1})}
\newcommand{\sect}[1]{Sect.\,\ref{#1}}
\newcommand{\tab}[1]{Table\,\ref{#1}}
\newcommand{\appx}[1]{Appendix\,\ref{#1}}

\documentclass{aa}  

\usepackage{xcolor}
\usepackage{graphicx}
\usepackage{txfonts}
\usepackage{mhchem}
\usepackage{environ}
\usepackage{array}
\usepackage{hyperref}
\hypersetup{
    colorlinks,
    citecolor=blue,
    filecolor=blue,
    linkcolor=blue,
    urlcolor=blue
}
\usepackage{datetime}

\usepackage{soul}

\definecolor{orange}{rgb}{1,0.5,0}
\definecolor{sred}{rgb}{.5,0,0}
\newcommand{\tg}[2]{{}{#2}}

\begin{document}

   \title{Mapping Synthetic Observations to Prestellar Core Models: \\ An Interpretable Machine Learning Approach}

   \author{T.~Grassi
          \inst{1,2}\fnmsep\thanks{Corresponding author,  \email{tgrassi@mpe.mpg.de}}
          \and
          M.~Padovani\inst{3}
          \and
          D.~Galli\inst{3}
          \and
          N.~Vaytet\inst{4}
          \and
          S.~S.~Jensen\inst{1}
          \and
          E.~Redaelli\inst{1,5}
          \and
          S.~Spezzano\inst{1}
          \and
          S.~Bovino\inst{3,6,7}
          \and
          P.~Caselli\inst{1}
          }

   \institute{ Max-Planck-Institut f\"ur Extraterrestrische Physik, Giessenbachstra{\ss}e 1, 85748 Garching, Germany 
         \and
             ORIGINS, DE
         \and
          INAF Osservatorio Astrofisico di Arcetri, Largo E. Fermi 5, 50125 Firenze, Italy
          \and
          European Spallation Source ERIC, Data Management and Software Centre, Asmussens Allé 305, DK-2800 Lyngby, Denmark
          \and
          European Southern Observatory, Karl-Schwarzschild-Stra{\ss}e 2, 85748 Garching, Germany 
          \and
          Chemistry Department, Sapienza University of Rome, P.le A. Moro, 00185 Roma, Italy
          \and
          Departamento de Astronom\'ia, Facultad Ciencias F\'isicas y Matem\'aticas, Universidad de Concepci\'on, Av. Esteban Iturra s/n Barrio Universitario, Casilla 160, Concepci\'on, Chile
             }

   \date{Received -; accepted -}

  \abstract
   {Observations of molecular lines are a key tool to determine the main physical properties of prestellar cores. However, not all the information is retained in the observational process or easily interpretable, especially when a larger number of physical properties and spectral features are involved.}
   {We present a methodology to link the information in the synthetic spectra with the actual information in the simulated models (i.e., their physical properties), in particular, to determine where the information resides in the spectra.}
   {We employ a 1D gravitational collapse model with advanced thermochemistry, from which we generate synthetic spectra. We then use neural network emulations and the SHapley Additive exPlanations (SHAP), a machine learning technique, to connect the models' properties to the specific spectral features.}
   {Thanks to interpretable machine learning, we find several correlations between synthetic lines and some of the key model parameters, such as the cosmic-ray ionization radial profile, the central density, or the abundance of various species, suggesting that most of the information is retained in the observational process.}
   {Our procedure can be generalized to similar scenarios to quantify the amount of information lost in the real observations. We also point out the limitations for future applicability.}

   \keywords{\dots
               }

   \maketitle
\section{Introduction}\label{sect:introduction}
Prestellar cores are cold and quiescent regions in molecular clouds representing the initial conditions for star formation \citep{Caselli2011}. They reveal complex chemical structures, with different molecular species tracing different properties of such objects (e.g., \citealt{Bergin2005, Spezzano2017}). This chemical variety makes it possible to employ some of the chemical species to infer physical properties, such as the object's geometry \citep{Tritsis2016}, temperature \citep{Crapsi2007,Harju2017,Pineda2022}, volume density \citep{Lin2022}, and gas kinematics \citep{Caselli2002,Punanova2018,Redaelli2022b}.

In addition to these properties, estimating the cosmic-ray ionization rate of \ce{H2} ($\zeta$) from observations remains a complex task, since each method carries some degree of uncertainty, mainly due to unknown rate coefficients in chemical networks and the source size. In a diffuse interstellar medium ($A_{\rm V}\leq1$\,mag), $\zeta$ can be inferred from \ce{H3+} absorption lines (e.g., \citealt{Indriolo2012}), and from estimating the gas volume density $n_{\rm H2}$ from 3D mapping of the dust extinction using Gaia data coupled with chemical modeling of \ce{C2} spectra \citep{Neufeld2024,Obolentseva2024}, the latter obtaining lower values of $\zeta$. At higher column densities, a wider number of methods have been developed using different molecular tracers (e.g., \citealt{Caselli1998,Bovino2020,Redaelli2024}). For a recent review, we refer the reader to \citet{Padovani2024}.

However, the exact dynamical model that best describes prestellar core evolution remains uncertain, with various models producing different predictions for gas molecular transitions, not only in prestellar cores (e.g., \citealt{Keto2008,Sipila2022}), but also in other objects, such as molecular clouds \citep{Priestley2023}. This suggests that determining the link between the features observed in the synthetic observations and the corresponding simulated objects is crucial to understanding prestellar cores' physical structure and dynamic evolution \citep{Jensen2023}.

Inferring physical properties from the features of observed spectra is a well-established methodology \citep{Kaufman1999,Pety2017}. However, only recent studies have explored the use of machine-learning techniques to predict gas parameters from observational data. For example, using random forests and multi-molecule line emissions, \citet{Gratier2021} inferred molecular clouds' H$_2$ column density from radio observations. \citet{Behrens2024} employed standard neural networks to determine the cosmic-ray ionization rate from observed spectra of HCN and HNC. In photon-dominated regions, \citet{Einig2024} found the most effective combination of tracers by comparing the mutual information of a physical parameter and different sets of lines using an approach based on conditional differential entropy. \citet{Shimajiri2023} used Extra Trees Regressor to predict the \ce{H2} column density from CO isotopologues. With explanatory regression model \citet{Diop2024} studied the dependence of CO spatial density on several protoplanetary disk parameters, including gas density and cosmic-ray ionization rate.

Finally, \citet{Heyl2023} \tg{}{and later \citet{AsensioRamos2024}} employed interpretable machine learning to determine how key physical parameters influence the observed molecular abundances. In particular, they assigned to each model's parameter feature a contribution value to a specific chemical species prediction using SHapley Additive exPlanations (SHAP, \citealt{Lundberg2017}), a technique based on cooperative game theory principles to interpret the predictions of a machine learning emulator. We will employ a similar approach, but in our case, we will invert the emulation process to predict model parameters from the observed spectra. Thanks to SHAP, we will determine the contribution of each spectral feature to each physical parameter.

In \sect{sect:steps}, we first give an overview of the different steps; \sect{sect:models} describes the procedure employed to generate the gravitational collapse models, their thermochemical evolution, and the post-processing with a large chemical network; \sect{sect:synth-obs} shows the details of the synthetic spectra calculations; In \sect{sect:emulators} we report the forward and backward emulation of the procedures described so far, while \sect{sect:shap} is dedicated to the interpretation of the results by using SHAP. Finally, \sect{sect:limitations} and \sect{sect:conclusions} report the limitations of our approach and the conclusions with future outlooks.

\section{Analysis steps}\label{sect:steps}
\begin{figure*}
    \centering
    \includegraphics[width=0.98\textwidth]{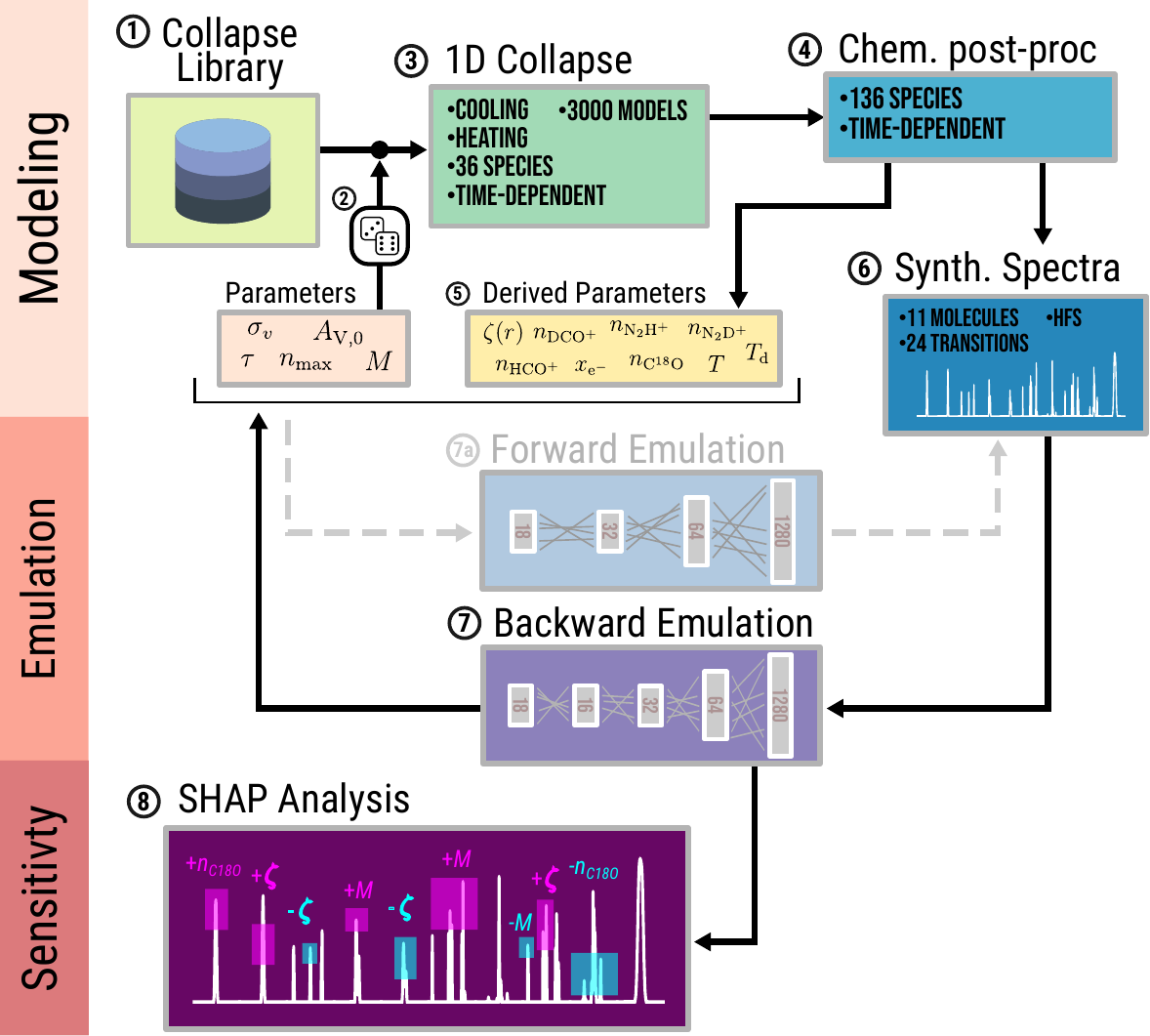}
    \caption{The procedure employed in this work. The top part represents the modeling steps: (1) Generate the library of gravitational collapse models, (2) randomly select the base models depending on the global parameters, (3) evolve the thermochemistry in the collapse model, (4) include additional chemistry with post-processing, (5) obtain the derived quantities, for example, the abundances of some key species, and (6) produce the synthetic spectra. The middle part is the emulation: (7a) is the forward emulation, from parameters to spectra, while (7) is the backward emulation, from spectra to parameters. The sensitivity using SHAP in the bottom is where we ``perturb'' the neural network input features to determine the impact on the outputs (8).} \label{fig:scheme}
\end{figure*}

To connect numerical models of prestellar cores and synthetic observations, we employ some of the recent advancements in interpretable machine learning. The general concept is illustrated in \fig{fig:scheme}. In particular, we generate a library of 1D hydrodynamical isothermal collapse models (label~1 in the figure). This provides the Lagrangian time-dependent tracer particles. We randomize a set of parameters to select and generate 3000 collapse models with different conditions that will evolve in time with a relatively small chemical network to determine the temperature evolution (2-3). Since we only use a simplified chemical network to determine the gas and dust temperature in the previous step, the following post-processing stage gives us a more complicated chemistry than the simple one needed for cooling and heating (4). We do not limit our analysis to the randomized parameters but also to derived quantities, like the cosmic-ray ionization radial profile or the abundances of the various molecules (5). Thanks to the additional chemical species, we will generate a database of synthetic spectra for several molecules and transitions (6). At this stage, we have a correspondence between initial parameters and spectra. This allows us to use a standard neural network to emulate the backward process (from spectra to parameters, 7) and, less relevant to the aims of this paper, the forward process (from parameters to spectra, 7a). In this way, we have a differentiable operator to determine the impact of the input (i.e., each velocity channel in the synthetic observations) onto the output (i.e., each physical parameter in the model). The weapon of choice for this analysis is the SHapley Additive exPlanations or SHAP (8).

\section{Model sample generation}\label{sect:models}

\subsection{Collapse model}\label{sect:hydro}
Before selecting the random parameters, we generate a library of collapse models for different initial masses ($M$). In this context, we assume that the temperature variations are negligible for the dynamics, but not for the chemistry. For this reason, we can generate the time-dependent trajectories of the gas elements in advance, assuming isothermality, and compute the gas and dust temperature in a second stage (\sect{sect:thermochem}).

To model the hydrodynamical collapse, we use the fully-implicit Godunov Lagrangian 1D code \textsc{Sinerghy1d}, which solves standard non-magnetized fluid equations \citep{Vaytet2012,Vaytet2013,Vaytet2017}. In this case, we assume the collapse to be isothermal, similar to the one described in \citet{Larson1969} and \citet{Penston1969}. Setting the mass $M$ of the collapsing region allows computing the collapsing radius $R_{\rm c}$ and the corresponding initial density $\rho_0$ as  
\beq
 R_{\rm c} = 0.41 \frac{G\,M}{c_{\rm s}^2}\qquad{\rm and}\qquad \rho_0 = \frac{M}{4/3\pi R_{\rm c}^3}\,,
\eeq
where $c_{\rm s}=\sqrt{k_{\rm B} T \mu^{-1}  m_{\rm p}^{-1}}$ is the isothermal speed of sound, $G$ the gravitational constant, $k_{\rm B}$ the Boltzmann constant, and $m_{\rm p}$ the proton mass. We assume constant temperature $T=10$\,K and mean molecular weight $\mu=2.34$, noticing that modifying these parameters within the ranges allowed by the collapse of a prestellar core plays a minor role in the evolution of the system. Considering that the gravitational pull on a volume element located at radius $r$ is proportional to $\int_0^{r} \rho(r')\,\dd r'$ (i.e., spherical self-gravity), we obtain a set of collapse profiles at different times. As expected, the collapse is self-similar. We note that the Larson collapse was discarded by \citet{Keto2015} for the special case of L1544. However, the \citet{Larson1969} model is easier to parametrize, and we use it here as an example. Other contraction models will be considered in future work.

To mimic the presence of magnetic pressure or similar physical processes that might reduce the collapse velocity, we introduce a unitless factor $\tau$ to slow down the collapse. This is conceptually equivalent to slowing each Lagrangian particle's ``clock'' by a factor $\tau$ (cf., e.g., \citealt{Kong2015,Bovino2021}). Therefore, the hydrodynamical timestep of each particle becomes $\Delta t = \tau \Delta t_\mathcal{H}$ and the corresponding radial velocity $\varv = \varv_{\mathcal{H}} / \tau$, where the subscript $\mathcal{H}$ indicates the quantity computed by the hydrodynamical code.

\subsection{Parameter generation}\label{sect:synthpop}
To generate one of the 3000 prestellar cores in our synthetic population sample, we perform the following steps:
\begin{enumerate}
\item Randomly select an initial mass $M$ in the range 5\,M$_\odot$ to 15\,M$_\odot$ and the corresponding collapse model calculated as described in \sect{sect:hydro};
\item Select a maximum central number density using a uniform random generator in the logarithmic space within a range of $10^5$\, cm$^{-3}$ and $10^7$\, cm$^{-3}$. This corresponds to a specific maximum time $t_{\rm max}$ that depends on the model selected in the first step; 
\item Randomize the visual extinction at the limits of the computational domain\footnote{The computational domain is $1.5\times10^5$\,au for each model, i.e., enough to ensure that every model has an external region unaffected by the collapse. In other words, where the initial conditions (density and velocity) are unperturbed.} ($A_{\rm V, 0}$) between 2 and 7\,mag. This determines the corresponding column density as $A_{\rm V, 0}=1.0638\times10^{-21} N_0$, where the multiplying factor is in units of mag\,cm$^{2}$. The aim is to simulate the presence of an external cloud environment;
\item Randomize the time factor $\tau$ uniformly between 1 and 10 and scaling timesteps, velocities, and adiabatic heating accordingly;
\item Randomize the radially uniform and non-time-dependent velocity dispersion of the turbulence between $r_{\sigma}=0.01$ and $r_{\sigma}=0.33$ times the maximum of the absolute value of the velocity during the collapse, i.e., $\sigma_\varv=r_\sigma \max(|\varv|)$;
\item Compute the time-dependent gas and dust temperature radial profile using a 275-reactions chemical network, as described in \sect{sect:thermochem};
\item Compute the time-dependent gas chemistry using gas and dust temperature from the previous step, but using a 4406-reactions chemical network, as described in \sect{sect:postproc};
\end{enumerate}

These steps are repeated for all the 3000 models using the aforementioned randomization.
This procedure will generate a radial profile of every quantity (e.g., density, chemical abundances, temperature, cosmic-ray ionization rate) for each model at each time step. Still, we will limit our analyses to $t=t_{\max}$.

An example model from the set of synthetic cores is shown in \fig{fig:model}, where the upper-left panel shows the total number density radial profile obtained with the hydrodynamical code at $t_{\rm max}$, while the upper-right panel reports the radial velocity profile. The temperature and the cooling and heating process in the upper-right and lower-left panels are discussed in \sect{sect:thermochem}, while the chemical abundances in the lower-right panel are discussed in \sect{sect:postproc}. The other models show similar radial profiles.

\subsection{Thermochemistry}\label{sect:thermochem}
In order to generate various temperature profiles, we use a simplified chemical network.
Since each Lagrangian particle has defined time, density, velocity, and position, we can compute its time-dependent thermochemistry. To this aim, we use the thermochemistry code \textsc{KROME} \citep{Grassi2014}, with a limited chemical network of 36~species (see \appx{sect:small-network}). We include 275 reactions from the \texttt{react\_COthin\_ice} network from \textsc{KROME}, mainly based on \citet{Glover2010} chemistry. More in detail, it employs gas chemistry, $A_{\rm v}$-based photochemistry, cosmic-ray chemistry, and CO and water evaporation and freeze-out. The network includes \ce{H2} \citep{Draine1996,Richings2014} and CO \citep{Visser2009} self-shielding parametrizations.  This set of reactions is optimized to produce the correct amount of coolant species and, therefore, the correct gas temperature (e.g., see \citealt{Glover2012,Gong2017}).

The initial conditions are the same for all the Lagrangian particles that, at time zero, have the same density by construction, as described in \sect{sect:hydro}. For this chemical network, the initial conditions are reported in \tab{tab:initial_conditions}, labeled as \emph{small} and \emph{both}. The chemical initial conditions play a major role in the time-dependent chemical evolution of several species, for example, HCN \citep{HilyBlant2010}. In this work, we arbitrarily decided to use the initial conditions of \citet{Sipila2018}, but other initial conditions could also be employed, e.g., \citet{Bovino2019}. Given the aims of the present work, we do not explore this issue further.

We use a radial-dependent cosmic-ray ionization rate model. To this aim, we employ the propagation model from \citet{Padovani2018} where the attenuation of the Galactic CR spectrum  is calculated considering the energy loss processes due to collisions between cosmic rays and hydrogen molecules. In particular, we used the $\alpha=-0.4$ model. The ionization rate of molecular hydrogen $\zeta_{\rm \ce{H2}}(r)$ is a function of the integrated column density at the given radius, $N(r) = N_0+\int_{R_{\rm c}}^r n(r')\,\dd r'$, where $N_0$ is the column density at the boundary of the computational domain, see \sect{sect:synthpop}. Since the column density is time-dependent, the cosmic-ray ionization rate also varies with time, impacting the chemistry and heating evolution.

We briefly describe the thermal processes employed in \textsc{KROME}, referring the reader to the latest repository commit\footnote{\url{https://bitbucket.org/tgrassi/krome}, commit \texttt{6a762de}.} and the references for additional technical details.

The radiative cooling is calculated for molecular hydrogen (with H, \ce{H+}, \ce{H2}, \ce{e-}, and \ce{He} colliders, see \citealt{Glover2015}), CO \citep{Omukai2010}, metal cooling \citep{Maio2007,Sellek2024}, atomic carbon (3 levels with H, \ce{H+}, \ce{e-}, and \ce{H2} colliders), atomic oxygen (3 levels with H, \ce{H+}, and \ce{e-} colliders), and \ce{C+} (2 levels with H and \ce{e-} colliders). \textsc{KROME} has its own dust cooling module, but for comparison, here we employ \citet{Sipila2018},
\beq
\Lambda_{\rm d} = 2\times10^{-33} n_{\ce{H2}}^2 (T - T_{\rm d}) \sqrt{\frac{T}{10\,{\rm K}}}\qquad{\rm erg\,cm^{-3}\,s^{-1}}\,,
\eeq
where the dust temperature $T_{\rm d}$ is determing by using the approach of \citet{Grassi2017}, assuming an MRN size distribution in the range $a_{\rm min}=5\times10^{-7}$\,cm  and $a_{\rm max}=2.5\times10^{-5}$\,cm, Draine's radiation \citep{Draine1978}, and optical constants from \citet{Draine2003}. We noted that different optical properties (for example \citealt{Ossenkopf1994}) impact the calculated temperature, but play a minor role in the aim of this paper. We assume a constant dust-to-gas mass ratio of 0.01 and use the same dust distribution to model the surface chemistry.
Additional processes that we included for completeness but that play a negligible role are \ce{H2} dissociation, Compton, and continuum cooling \citep{Cen1992}.

The thermal balance is completed with photoelectric heating \citep{Bakes1994,Wolfire2003}, photoheating (in particular \ce{H2} photodissociation and photopumping), cosmic-ray heating following \citet{Glassgold2012} and \citet{Galli2015}, heating from \ce{H2} formation, and compressional heating $n_{\rm tot} k_{\rm B} T \,t_{\rm ff}^{-1}$, where the free-fall time is $t_{\rm ff}=\sqrt{3 \pi / \left(32\,  G \mu m_{\rm p} n_{\rm tot}\right)}$.

Although we are employing a time-dependent thermochemistry model on Lagrangian particles computed with isothermal hydrodynamics, we noted that the small range of computed temperatures has little or no impact on the collapse model. In addition, this uncertainty is already superseded by other more important uncertainties, such as the inclusion of the time factor $\tau$. To assess the validity of our code, we benchmarked our result with a set-up similar to \citet{Sipila2018}, obtaining a good agreement between their results and ours. Despite using the same initial conditions and similar physics for this specific comparison (e.g., constant cosmic-ray ionization rate), the differences are due to different chemistry and cooling mechanisms, such as a different CO cooling method. When we use the variable cosmic-ray, the temperature profile shows additional discrepancies from \citet{Sipila2018} due to a different heating prescription and radial variability. We have, in general, smaller dust temperature values due to the different grains' optical properties (more details in \appx{sect:comparison}). Despite these differences, the main findings and the general method described in this work are not influenced. We aim to address the role of the uncertainties in a future paper.

\begin{table}
    \centering
    \begin{tabular}{lll}
        \hline
        Species      &  Fractional abundance & Network \\
        \hline
        p-\ce{H2} & see text & large\\
        o-\ce{H2} & see text & large\\
        \ce{H2} & 0.5 &  small\\
        \ce{HD} & 1.60(-5) & large\\
        \ce{He} & 9.00(-2) & both\\
        \ce{O} & 2.56(-4) & both\\
        \ce{C+} & 1.20(-4) & both\\
        \ce{N} & 7.65(-5) & large\\
        \ce{e-} & Same as \ce{C+} & both\\
        Default & 0.00 & both\\
        \hline
    \end{tabular}
\caption{Chemical initial conditions specific for the \emph{small} chemical network of \sect{sect:thermochem}, for the \emph{large} chemical network of \sect{sect:postproc}, or for \emph{both} \citep{Sipila2018}. Fractional abundances are with respect to the total abundances inferred from the mass density obtained from the hydrodynamical code assuming the constant mean molecular weight $\mu$. Their format is ${a(b)=a\times10^b}$.}\label{tab:initial_conditions}
\end{table}

In the upper-right panel of \fig{fig:model}, we report the gas and dust temperature radial profiles, while in the lower-left panel, we report the main thermochemical quantities for an example model of our synthetic population sample.

\subsection{Chemical postprocessing}\label{sect:postproc}
In this work, we are interested in molecules such as \ce{DCO+}, o-\ce{H2D+}, \ce{N2H+}, \ce{N2D+}. The modeling of their evolution requires a more comprehensive network than the one in \sect{sect:thermochem}. To this aim, we employed the network described in \citet{Bovino2020} that includes 4406~reactions and 136~chemical species (see \appx{sect:large-network}).

The chemical initial conditions employed for this network are reported in \tab{tab:initial_conditions}, labeled as \emph{large} and \emph{both}. For \ce{H2}, we assume an initial fractional abundance of 0.5 and an ortho-to-para ratio of $10^{-3}$ \citep{Pagani2013,Lupi2021}. As mentioned in the previous section, the initial conditions are arbitrary and play a major role in the chemical evolution of the prestellar model. However, for the aims of the present paper, we do not address this issue.

This post-processing approach allows the evolution of the complex network to be locally independent of temperature evolution. In other words, the Lagrangian particle evolution temperature is time-dependent (as computed in \sect{sect:thermochem}), but we assume the chemical evolution to be at constant temperature during each timestep. This allows a much faster calculation since, at each call, the solver's Jacobian is not influenced by temperature variations, greatly reducing the stiffness of the ordinary differential equation system. In principle, we could use \textsc{KROME} to evolve the temperature together with the 4406-reactions network, but this will greatly increase the computational time and create potential numerical instabilities. At the present stage, for development efficiency reasons, we prefer to focus on reducing the computational cost rather than achieving full consistency. We plan to improve the pipeline in future works.

We verified the correctness of our approach by comparing the abundance of the coolant and the colliders obtained with the two chemical networks, finding no relevant differences in the cooling functions.

\begin{figure*}
    \centering
    \includegraphics[width=0.98\textwidth]{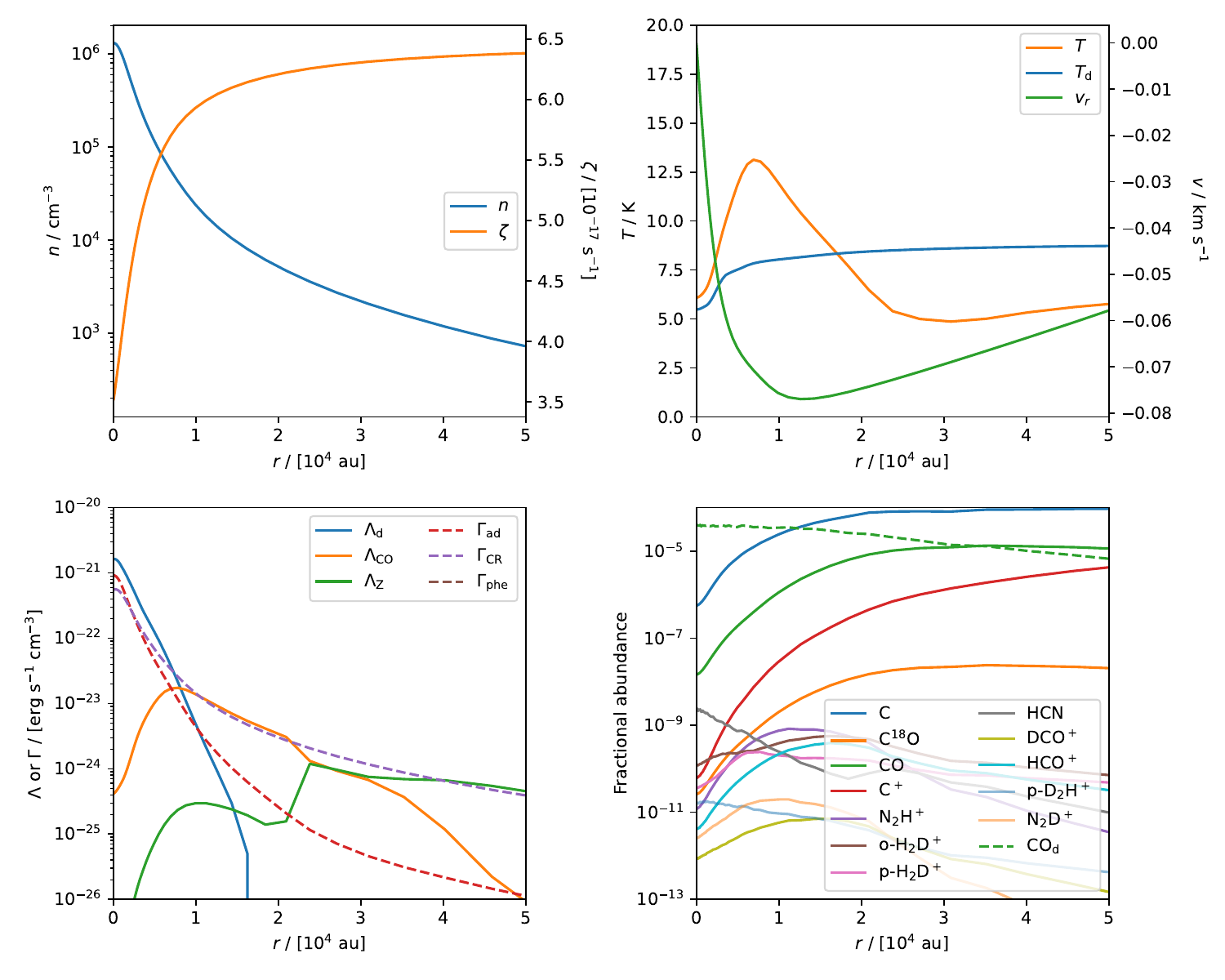}
    \caption{Example model from the population synthesis set at $t_{\rm max}$. The $x$-axis of each panel spans approximately 0.3\,pc, or 350" assuming a cloud distant 170\,pc from the observer. For the sake of clarity, this plot shows a smaller inner region of the actual computational domain. Upper left panel: total number density radial profile (blue, leftmost $y$-scale), cosmic-ray ionization rate $\zeta$ (orange, rightmost $y$-scale). Upper right: gas (orange) and dust (blue) temperature radial profiles (both on leftmost $y$-scale), and radial velocity profile (green, rightmost $y$-scale). Lower left: Cooling and heating contributions, in particular $\Lambda_{\rm d}$ is the dust cooling, $\Lambda_{\ce{CO}}$ is the \ce{CO} cooling, $\Lambda_{\rm Z}$ is the cooling from atomic species (C, \ce{C+}, and O), while $\Gamma_{\rm ad}$ is the adiabatic heating (i.e., compressional heating), $\Gamma_{\rm CR}$ the cosmic-ray heating, and $\Gamma_{\rm phe}$ the dust photoelectric heating. The cooling and heating contributions below $10^{-26}$\,erg\,s$^{-1}$\,cm$^{-3}$ are not reported in the legend. Lower right:  the radial profile of a subset of the chemical species computed with the 4406-reactions network. The fractional abundance is relative to the total number density. CO$_{\rm d}$ (dashed green) is the CO on the dust surface. The comparison with \citet{Sipila2018} is reported in \appx{sect:comparison}.} \label{fig:model}
\end{figure*}

In the lower-right panel of \fig{fig:model}, we show an example of some selected chemical species radial profiles, including the tracers employed later in the synthetic observations.

\subsection{Model parameter definition and their correlation}\label{sect:parameters}
In \sect{sect:synthpop}, we defined some randomly sampled parameters: maximum central number density ($n_{\rm max}$), visual extinction at the limit of the computational domain ($A_{\rm V,0}$), turbulence velocity dispersion ($\sigma_\varv$), collapse time factor ($\tau$), and the mass of simulation domain ($M$). To infer additional information from our models, we define some derived parameters\footnote{Specifically, these are not ``parameters'' since they are derived quantities, but for the sake of simplicity, we refer to them as parameters anyway.}. These are reported in \tab{tab:parameters} and evaluated at a specific arbitrary distance of $10^4$\,au.
In addition, we fit the radial profile of the cosmic-ray ionization rate at $t=t_{\rm max}$ and for $r\leq4\times10^4$\,au. We employ a four-parameter dimensionless function 
\beq\label{eqn:crfit}
 \hat{\zeta}(r') = \frac{a_0}{1+\exp\left[a_1(r' - a_2)\right]} + a_3\,,
\eeq
where $r' = 0.57\,\left[\log(r / \rm{au}) - 1\right] - 1.2$ and $\hat{\zeta} = \zeta / (10^{-17}\,{\rm s}^{-1}) - 4$. The additional parameters are therefore $a_0$, $a_1$, $a_2$, and $a_3$. Despite the change of coordinates, $\hat{\zeta}(r')$ has the same physical interpretation and properties of $\zeta(r)$. This relation is empirically determined \emph{a posteriori} from the data and has no intentional physical meaning.

{\setlength{\extrarowheight}{4pt}
\begin{table}
    \centering
    \begin{tabular}{llll}
        \hline
        Var. & Rnd. & Description & Units\\
        \hline
        $n_{\rm max}$ & log & Max central density & cm$^{-3}$\\
        $\sigma_\varv$ & lin & Turbulence velocity dispersion & km\,s$^{-1}$\\
        $A_{\rm V,0}$ & lin & $A_{\rm V}$ at domain boundary & mag\\
        $\tau$ & lin & Collapse time factor & -\\
        $M$ & lin & Total mass & $M_\odot$\\
        $T^{1e4}$ & - & $T$ at $r=10^4$\,au and $t=t_{\rm max}$ & K\\
        $T^{1e4}_{\rm d}$ & - & Same as $T^{1e4}$ but for $T_{\rm d}$ & K\\
        $n_{\ce{C^18O}}^{1e4}$ & - & Same as $T$ but for $n_{\ce{C^18O}}$ & cm$^{-3}$\\
        $n_{\ce{HCO+}}^{1e4}$ & - & Same as $T$ but for $n_{\ce{HCO+}}$ & cm$^{-3}$\\
        $n_{\ce{DCO+}}^{1e4}$ & - & Same as $T$ but for $n_{\ce{DCO+}}$ & cm$^{-3}$\\
        $n_{\ce{N2H+}}^{1e4}$ & - & Same as $T$ but for $n_{\ce{N2H+}}$ & cm$^{-3}$\\
        $n_{\ce{N2D+}}^{1e4}$ & - & Same as $T$ but for $n_{\ce{N2D+}}$ & cm$^{-3}$\\
        $x_{\ce{e-}}^{1e4}$ & - & Same as $T$ but for $x_{\ce{e-}}$ & -\\
        $\zeta^{1e4}$ & - & Same $T$ but for $\zeta$ & s$^{-1}$\\
        $a_i$ & - & Coefficients of \eqn{eqn:crfit} $i\in[0,3]$ & -\\
        \hline
    \end{tabular}
\caption{Parameters employed in this work. Rnd. is the randomization distribution, not employed for the derived parameters. Note that for the electrons, we employ their fraction rather than their number density.}\label{tab:parameters}
\end{table}
}

\begin{figure*}
    \centering
    \includegraphics[width=1.1\textwidth]{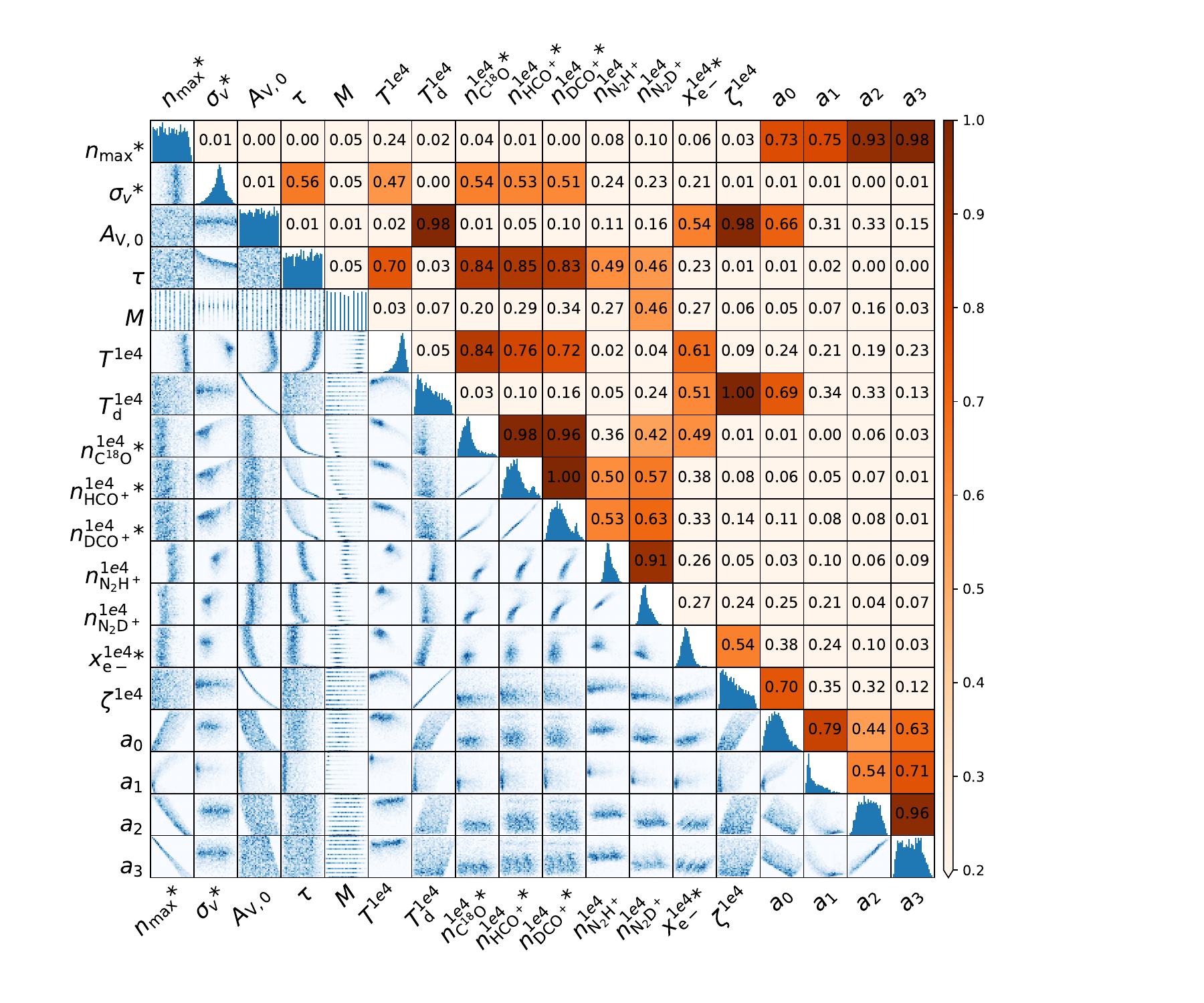}
    \caption{Correlation matrix between models' parameters. The panels in the upper left triangle show the Pearson correlation coefficient color-coded, as indicated by the color bar. The plots in the lower right triangle show the 2D correlation histograms of the parameters of the 3000 generated models. The panels in the diagonal represent the histogram of the parameter distribution. The correlation for quantities marked with "*" is computed using their logarithm. Note that the color bar is clipped to 0.2 to enhance the stronger correlations. \label{fig:correlation}}
\end{figure*}

\fig{fig:correlation} reports the correlations and the distribution of the sampled parameters in the 3000 models. The asterisk (${}^*$) in the superscript indicates a logarithmic sample; otherwise, it is assumed to be linear. The sampling type is manually chosen to maximize the uniformity of the distribution.
The panels on the diagonal of \fig{fig:correlation} show the distribution of the selected parameters.
As expected, the randomized parameters are uniformly sampled, apart from $\sigma_\varv$, which is a function of the maximum radial velocity and, therefore, shows a non-uniform distribution.

Conversely, the derived parameters show some degree of correlation with the random parameters and/or between them, as they are constrained by physics. The two-dimensional histograms in the lower triangle of the matrix of panels indicate their sampling of the parameter space and their correlation. These plots already show that there are some ``forbidden'' combinations, for example, large dust temperature ($T_{\rm d}^{1e4}$) and large visual extinction ($A_{\rm V, 0}$) at the same time. Analogously, in the upper triangle, we report the absolute value of the Pearson correlation algorithm coefficient for pair parameters to measure the degree of correlation. For example, $A_{\rm V, 0}$ and $T_{\rm d}^{1e4}$ strongly correlate since $A_{\rm V}$ is a key factor to determine the impinging radiation on a dust grain. For the same reason, the dust temperature correlates with cosmic-ray ionization rate since $\zeta$ is scaled with the external column density, which is also proportional to  $A_{\rm V, 0}$. The time factor $\tau$ correlates with the chemical species with long chemical time scales, which also correlates with $\sigma_\varv$, being a fraction of the maximum radial velocity that is scaled by $\tau$.

It is important to note that these relations are calculated at $10^4$\,au, hence, any radial effect is neglected. The dependence with the radius is included in the $a_i$ coefficients. As expected, they correlate with each other, and in particular, $a_0$, being the scaling factor of \eqn{eqn:crfit}, correlates with $\zeta^{1e4}$.

Despite most of the parameters having no radial dependence, this is implicitly present in the synthetic observations of the key chemical species, which, by construction, include information from different parts of the observed object.

\section{Synthetic observations}\label{sect:synth-obs}
Since we have the radial profile of the chemical composition of each model, we can produce synthetic spectral observations to mimic an observed prestellar core at time $t_{\rm max}$. 
We then construct a spherical core based on the specific radial profile obtained by the chemical evolution. In this way, we have a 3D structure with chemical abundances, velocity, and temperature. This allows us to employ the publicly available code \textsc{LOC}\footnote{\url{https://github.com/mjuvela/LOC} commit \texttt{643624b}.} which take into account the aforementioned quantities and the turbulence velocity dispersion. LOC is an OpenCL-based tool for computing the radiative transfer modeling of molecular lines \citep{Juvela1997,Juvela2020}.

We designed a pipeline to automatically produce synthetic spectra for each model using the GPU implementation of the code. 
The molecular transitions considered are between approximately 0.2 and 4\,mm as reported in \tab{tab:transitions}, alongside their main characteristics and references. The data is further described in the EMAA\footnote{\url{https://emaa.osug.fr/species-list}} and LAMDA\footnote{\url{https://home.strw.leidenuniv.nl/~moldata/}} databases. Since the large chemical network described in \sect{sect:postproc} does not include any oxygen isotopologues, we assume that the abundance of \ce{C^18O} is scaled from the \ce{CO} abundance by using a constant factor $1/560$ \citep{Wilson1994,Sipila2022}. We are aware that this approach is a simplification, and we will expand the chemical network in a future version of the code by including more isotopologues. As implemented, we expect any information we obtain on \ce{C^18O} to be a property of CO. However, this value is supposed to be a valid assumption, at least in the local ISM, and, in addition, the expected oxygen fractionation is modest \citep{Loison2019}.

We use the recently modified version of LOC, which allows us to include the radial profile of the main species and the radial profiles of the various colliders. Our tests show that these recent changes do not significantly affect the results, especially for molecules with molecular hydrogen as the main collider. 

We arbitrarily assumed a distance from the observer of 170\,pc for all the models, similarly to real observed objects (e.g., in the Taurus Molecular Cloud Complex \citealt{Torres2007,Lombardi2008,Galli2019}). For all the transitions, we employ a bandwidth of 4\,km\,s$^{-1}$ and 128 channels, except for the non-LTE hyperfine structure calculations of \ce{N2H+}, \ce{N2D+}, and \ce{HCN} where the bandwidth is incremented internally by LOC to accommodate all the observable lines. In this case, the spectra have been later interpolated over a 128-channel grid in the range of the new bandwidth, obtaining a smaller velocity resolution for these molecules. The interpolation does not show any significant distortion from the original spectrum, and this remapping is not relevant to the aims of this work.

\textsc{LOC} produces a 2D map of the source for each velocity channel. The spectra are then convoluted with a 2D Gaussian function to mimic a telescope beam as in \tab{tab:transitions}. To this aim, we use a customized version of the LOC convolution algorithm that produces the same results as LOC, but it is more efficient \emph{for our setup}. After this step, we extract the spectrum in the center of the source, obtaining a 1D function of the velocity channels.

\begin{figure*}
    \centering
    \includegraphics[width=0.98\textwidth]{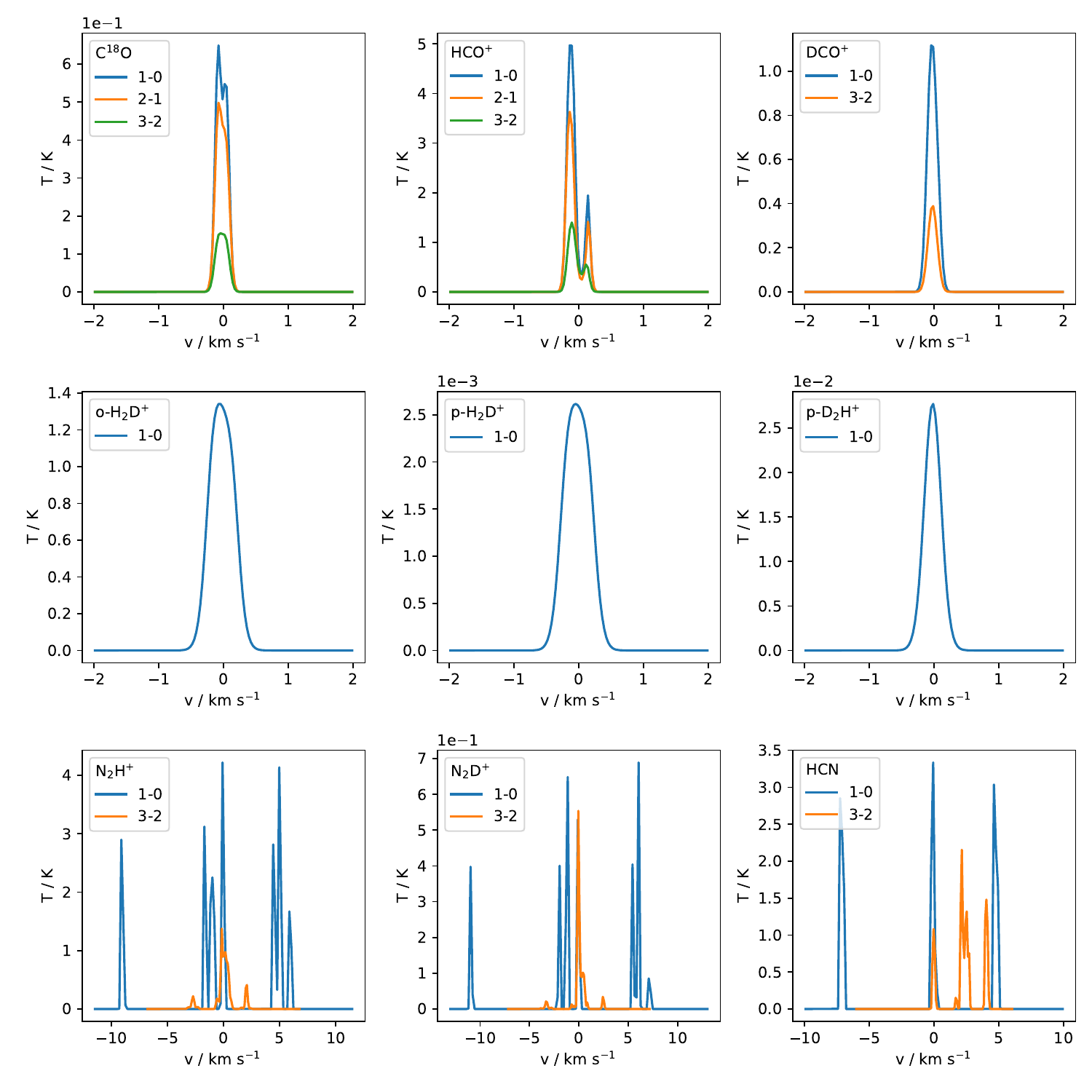}
    \caption{Example spectra for the model in \fig{fig:model}. Each panel reports the spectra of the molecule and the transitions indicated in the legend. The spectra are convoluted with a telescope beam, as discussed in the main text. Note that the $y$-axis is scaled to the value on the top of the panel, e.g., \ce{C^18O} temperature is scaled to $10^{-1}$. All the spectra are calculated with a $-2$ to $+2$\,km\,s$^{-1}$ bandwidth range and 128~channels, except for \ce{N2H+}, \ce{N2D+}, and HCN where the bandwidth is calculated by \textsc{LOC} to take into account the hyperfine structure of these molecules, but interpolated to use 128~channels. We also use the spectra centered according to the LOC output (i.e., the position of 0\,km\,s$^{-1}$).} \label{fig:spectra}
\end{figure*}

In \fig{fig:spectra}, we report an example of one of the calculated synthetic spectra. The solid lines represent the convoluted spectra after being produced by \textsc{LOC}. Note again that since \ce{N2H+}, \ce{N2D+}, and \ce{HCN} are modeled with their non-LTE hyperfine structure, LOC accommodates these lines using a different velocity range than the default (-2 to 2 km\,s$^{-1}$). At this stage, we do not include any instrumental noise, hence the information is degraded only by the assumed telescope beam. We will add some noise in the next step.

\begin{table*}
    \centering
    \begin{tabular}{lllllllll}
        \hline
        Species & Transitions & Label & Beams size &  HFS & Coll. & Ref. & DB & Training\\
        \hline
        \ce{C^{18}O}    &$1-0$, $2-1$, $3-2$ & 1-0, 2-1, 3-2 & 22", 28", 28" & n & o/p-\ce{H2}& (1) & L& -\\
        \ce{HCO+}       & $1-0$, $2-1$, $3-2$ & 1-0, 2-1, 3-2 & 28", 28", 23" & n & o/p-\ce{H2} & (2) & L & 1-0, 3-2\\
        \ce{DCO+}       & $1-0$, $3-2$ & 1-0, 3-2& 36", 12" & n & \ce{H2} & (2) & L & 1-0, 3-2\\
        o-\ce{H2D+}     & $1_{10}-1_{11}$ & 1-0 & 16" & n & o/p-\ce{H2} & (3, 4) & L & 1-0\\
        p-\ce{H2D+}     & $1_{01}-0_{00}$& 1-0 & 28" & n & o/p-\ce{H2} & (3, 4) & L & -\\
        p-\ce{D2H+}     & $1_{10}-1_{01}$ & 1-0 & 28"  & n & o/p-\ce{H2} & (3) & L& -\\
        \ce{N2H+}       & $1-0$, $3-2$ & 1-0, 3-2& 27", 12" & y & p-\ce{H2} & (5) & E & 1-0, 3-2\\
        \ce{N2D+}       & $1-0$, $3-2$ & 1-0, 3-2 & 34", 12" & y & p-\ce{H2} & (5) & E & 1-0, 3-2\\
        \ce{HCN}        & $1-0$, $3-2$ & 1-0, 3-2 & all 28" & y & p-\ce{H2}, \ce{e-} & (6, 7) & E & 1-0\\ 
        \hline
    \end{tabular}
\caption{Molecules employed in this work and their transitions. The transitions column reports the quantum numbers $J$ or the $J_{K_a\, K_c}$.  The label column reports the labels in the LOC notation that we use for convenience in the discussion and in the plots. Each beam size corresponds to the transitions in the second column. HFS indicates if a non-LTE hyperfine structure calculation has been made, while Coll. indicates the collisional partners: \ce{H2} represents the sum of the abundances of o-\ce{H2} and p-\ce{H2}, while o/p-\ce{H2} indicates that the collisional rates are available for both the colliders. LOC allows us to use the actual radial profile of the o/p-\ce{H2} ratio as computed by the chemical code. References are (1) \citet{Yang2010}, (2) \citet{DenisAlpizar2020}, (3) \citet{Hugo2009}, (4) \citet{Koumpia2020}, (5) \citet{Muller2005}, (6) \citet{Faure2007}, and (7) \citet{Dumouchel2010}. DB indicates if the data are in the LAMDA (L) or EMAA (E) database. The transitions used for the NN training and in the analysis are listed in the last column. See the text for the details on the selected transitions.}\label{tab:transitions}
\end{table*}

\section{Backward emulation: from observations to models}\label{sect:emulators}
The pipeline described so far can be interpreted as a ``forward'' operator $\mathcal{F}$ that produces spectra $\mathbf{s}$ from a set of parameters $\mathbf{p}$, i.e., $\mathbf{s} = \mathcal{F}(\mathbf{p})$. In our case, evaluating $\mathcal{F}$ takes a few minutes per parameters grid point. To reduce the computational cost, emulators allow mimicking the behavior of complicated (thermo)chemical systems, greatly reducing the time spent in evaluating the output (e.g., see \citealt{Grassi2011,Holdship2021,Grassi2022,SmirnovPinchukov2022, Palud2023,Sulzer2023,AsensioRamos2024,Branca2024,Maes2024}). An additional and relevant advantage of the reduced computational cost is the capability to quickly evaluate how input perturbations propagate to the output solution \citep{Heyl2023}. 

However, rather than evaluating $\mathcal{F}$, our aim is to understand how the information is degraded in this process, and in particular, how $\mathbf{s}$ can be employed to reconstruct $\mathbf{p}$. We therefore need to design the inverse (``backward'') operator $\mathcal{B} = \mathcal{F}^{-1}$, so that $\mathbf{p} = \mathcal{B}(\mathbf{s})$. Analogously to the forward case, we take advantage of emulation to invert the problem and later analyze how variations of $\mathbf{s}$ have an effect on $\mathbf{p}$.

Despite there being some established techniques to reconstruct some of the core parameters from some specific spectral information (for example, determining $\zeta$ from \ce{HCO+}, \ce{DCO+}, and other chemical tracers, as described in \citealt{Caselli1998} or  \citealt{Bovino2020}), in our case, we will use a blind approach that employs all the spectra, all the channels, and all the parameters at the same time. In particular, we will emulate the backward operator with a neural network (NN) that will ``learn'' to predict $\mathbf{p}$ from $\mathbf{s}$.

Following a standard machine learning procedure, we divide our set of 3000 models into a training (2100 models), a validation (450), and a test set (450). The training set is used to train the NN, the validation set is used to determine if we are \tg{oversampling}{overfitting} or \tg{undersampling}{underfitting} our data while training, and the test set is used to verify our predictions on ``new'' data. We visually verified that the distributions of each parameter in the sets are relatively uniform without performing any specific analysis, for example, to verify if some models in the test set are outside the convex hull of the training and validation set input features \citep{Yousefzadeh2021}.

We normalize each transition (i.e., the input data) in the ${[-1, 1]}$ range between zero and the maximum temperature of the whole set. For example, the transition 1-0 of \ce{HCO+} is normalized considering the maximum value of the given transition in all the training, validation, and test set models. This might overestimate the transitions with small maximum values (e.g., p-\ce{H2D+}). To avoid this, we introduce some Gaussian noise with a 10\,mK dispersion, eliminating all the information below this temperature threshold. We use the same noise for each molecule to avoid the NN recognizing the level of noise rather than the actual features of the emitted lines.

To reduce the computational impact, in particular the memory usage, we remove some of the transitions that we know are less relevant to our problem (e.g., the p-\ce{H2D+} signal is smaller than the noise we introduced). We remove \ce{C^18O} being its abundance a scaling of CO, but also as a test to determine if the remaining transitions are capable of inferring \ce{C^18O} and CO information even without the specific molecular lines. We also remove some additional transitions to simplify the final interpretation output, for example, \ce{HCO+} (2-1). The final lines are \ce{DCO+} (1-0, 3-2), HCN (1-0), \ce{HCO+} (1-0, 3-2), \ce{N2D+} (1-0, 3-2), \ce{N2H+} (1-0, 3-2), and o-\ce{H2D+} (1-0), as shown in the last column of \tab{tab:transitions}. This corresponds to 1280 velocity channels, i.e., 1280 input features in the NN.

We normalize the parameters in \tab{tab:parameters} (i.e., the output data) in the ${[-1, 1]}$ range, using the logarithmic values of $n_{\rm max}$, $\sigma_\varv$, and  the abundances at $r=10^4$\,au of \ce{C^18O}, \ce{HCO+}, and \ce{DCO+}, while the actual value for the other parameters. This choice maximizes the uniformity of the distribution. We have 18 parameters, i.e., 18 output features in the NN.

The NN is composed of 3 fully connected hidden layers of 64, 32, and 16 neurons each. The activation between each layer is a ReLU function, apart from the last two (i.e., the last hidden layer and output layer), which are directly connected. The code is implemented in Pytorch\footnote{Version 2.0.0.}, using an MSE loss function and an Adam optimizer with a learning rate of $10^{-3}$. All the other parameters follow the Pytorch default.

\tg{}{The NN architecture is designed to mimic an (auto)encoder in order to have a rough estimate of the information compression via dimensionality reduction \citep{Ballard1987,Champion2019,Grassi2022}. The specific number of dimensions is empirically determined by verifying the test set predictions while reducing the number of hidden neurons at every training attempt.} We do not use any hyper-parameter optimization \citep{Yu2020}.

\tg{}{The number of neurons in the first hidden layer is compatible with the PCA\footnote{Using \texttt{PCA} from \texttt{sklearn.decomposition} with default options, version 1.0.2. \citep{Halko2011,Pedregosa2011}.} of the input data set, where 99.99\% of the variance is explained by 69 components (cf.~\citealt{Palud2023}). Analogous consideration can be made for the last hidden layer, which is smaller than the output layer. Even in this case, increasing the size of the last hidden layer does not affect our results.}

\begin{figure}
    \centering
    \includegraphics[width=0.48\textwidth]{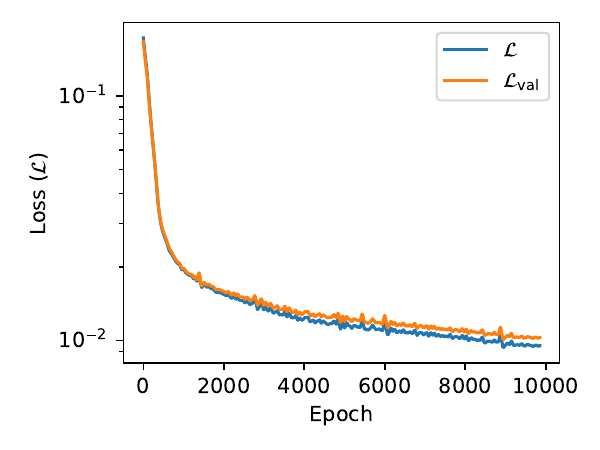}
    \caption{Training (blue) and validation (orange) total loss at different epochs for the backward emulation training. \tg{}{Note that although the total loss does not become constant, the individual losses of the physical parameters (not reported here) become constant after a few epochs, apart from the turbulence velocity dispersion ($\sigma_\varv$) and the total mass ($M$), that drive the total loss.}} \label{fig:loss}
\end{figure}

We stop the training after \tg{5000}{$10^4$} epochs to avoid overfitting, as shown by the total training and validation losses in \fig{fig:loss}. The noise in both losses is produced by the individual losses of some specific parameters, like the total mass ($M$) or the dust temperature at $10^4$\,au ($T_{\rm d}^{1e4}$). \tg{}{Analogusly, the turbulence velocity dispersion ($\sigma_\varv$) and the total mass ($M$) prevent the total loss from becoming constant (while we have verified that all the other individual losses become constant after a few thousand epochs).} Some strategies could improve this behavior: we can use an NN for each physical parameter, add weights in the MSE terms to reduce or increase the efficiency of the ``correction'' of a specific parameter, or change the parameters of the optimizer (e.g., the learning rate). However, at this stage, we do not optimize the training to keep the method as generic as possible.

\begin{figure*}
    \centering
    \includegraphics[width=0.98\textwidth]{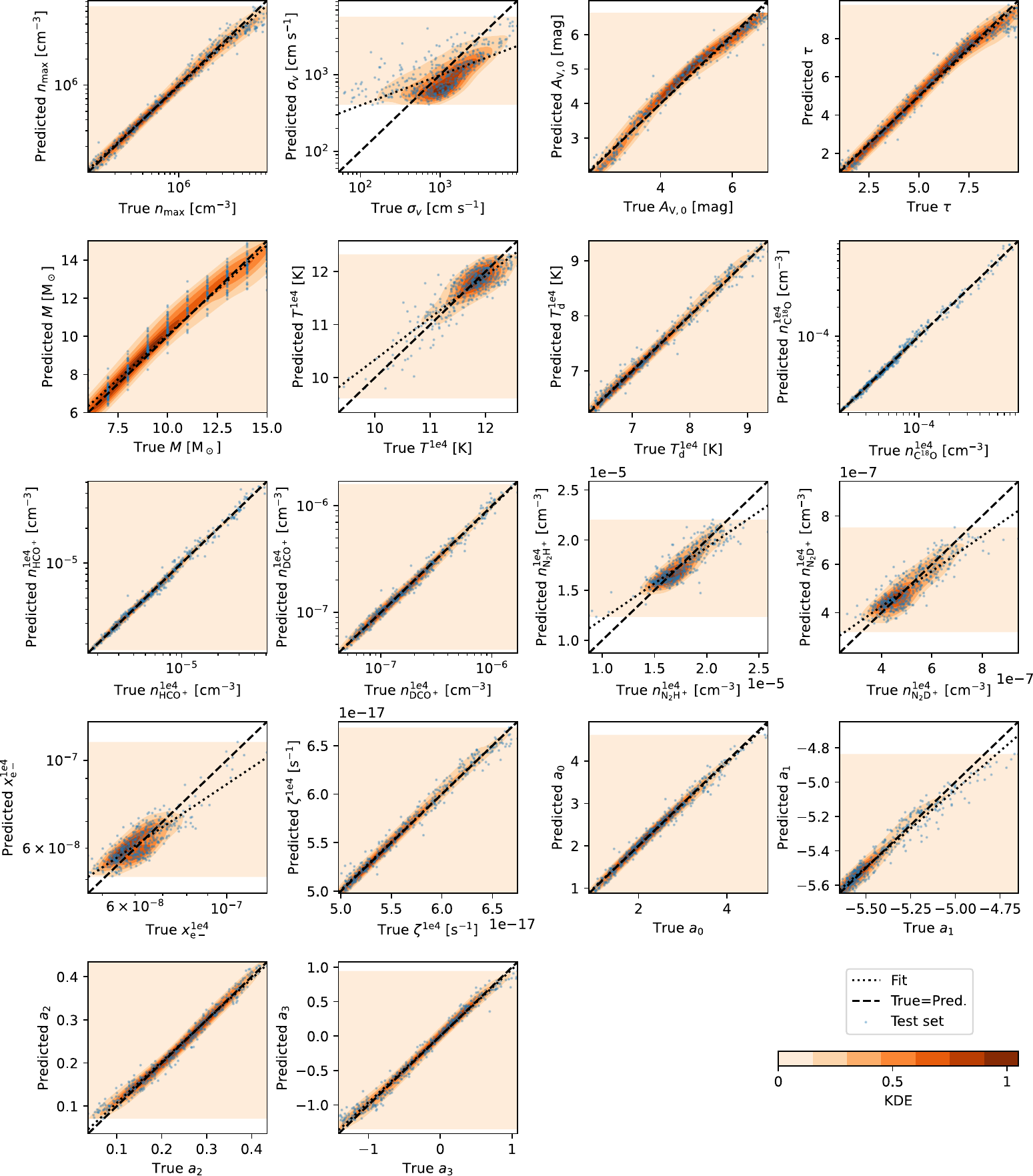}
    \caption{Comparison between the true ($x$-axis) and the predicted ($y$-axis) parameter values for the models in the test set (scatter plot points). Each panel is calculated for a specific model parameter (see axis labels). A perfect match between true and predicted values corresponds to the dashed line. The dotted line represents the fit of the scatter points, while the color contours are their Kernel Density Estimation (KDE) normalized to the maximum value. The KDE color scale in the last panel is the same for all the panels. Note that $\langle \zeta \rangle$ is in units of $10^{-17}$~s$^{-1}$.} \label{fig:kde}
\end{figure*}

To better assess the error for each parameter, we compare the predictions of the NN using the models in the test set. The result is the scatter plot in \fig{fig:kde}. The dashed line in each panel represents the perfect match between true and predicted values, while the dotted line is the fit of the scatter plot points. Ideally, all the points should lie on the dashed line. We added the contours calculated using a Gaussian Kernel Density Estimation (KDE) algorithm (using the Scott method for the bandwidth estimator, \citealt{Scott1992}) to provide a statistical overview of the results. The KDE represents the probability distribution of the data points.

We note that most quantities are well-reproduced by the NN, apart from $\sigma_\varv$ and $M$. This indicates that the spectra retain the information on most of the model parameters. As a proof of concept, we applied the same procedure by adding a Gaussian noise of 5\,K to the spectra (i.e., comparable to the highest intensities found). In this case, the NN fails to reproduce every parameter. For a similar reason, as shown by the $\sigma_\varv$ panel in \fig{fig:kde}, the NN fails to predict very small velocity dispersions, i.e., it cannot determine $\sigma_\varv$ below a certain threshold.

It is worth noticing that although the NN fails to reproduce $\sigma_\varv$ and $M$, this does not necessarily mean that the spectra do not retain the information on these parameters. In fact, the reason could be a standard NN failure in learning the connection between input and output (see the discussion on the loss above). \tg{}{This issue could be further explored by the use of mutual information as discussed by \citet{Einig2024}.} On the other hand, an accurate NN parameter reproduction indicates that the spectra actually retain the information (sufficient condition).

\tg{}{To quantify the error, we follow a procedure similar to \citet{Palud2023} by comparing for each parameter the statistical properties of the relative error distribution in the linear space. By calculating the median and the maximum from each error distribution, we found that the medians of all the quantities are below 5\%, apart from $n_{\rm max}$, $n_{\rm N2D+}^{1e4}$, and $a_2$, that are around 10\%, and $a_3$ and $\sigma_{\varv}$ that are around the 100\%. The best-reproduced quantity is $\zeta^{1e4}$ with a median around 0.7\%. The maximum error in some cases reaches 50\% (the molecular abundances, $\tau$, and $n_{\rm max}$), while it is generally around the 20\%, if we exclude again $a_3$ and $\sigma_{\varv}$. However, it is important to notice that a single isolated outlier test model causes the maximum error, and the 90th percentile of each error distribution is always well below this maximum error. Furthermore, although the relative error gives a good estimate of the NN performance, we remark that the accuracy depends on each quantity's physical meaning and the desired precision for the specific scientific problem.}
\tg{In addition}{In our case, for example}, to verify the NN prediction on $a_i$, we compare the predicted and expected \eqn{eqn:crfit} for the models in the test set, finding a negligible error \tg{}{(well below $5\times10^{-18}$\,s$^{-1}$)}, as discussed in more detail in \appx{sect:cr-fit}. \tg{}{Therefore, in an ideal set-up, the NN loss for the $a_i$ values should be defined on the precision required by \eqn{eqn:crfit}.}

The analysis presented in this section only tells us whether the parameters are well-reproduced, while in \sect{sect:shap}, by analyzing the corresponding spectra, we will investigate which emission line plays a role in determining the final parameter values. \tg{}{The forward emulator is discussed in \appx{sect:forward}}.

\section{Results of the SHAP Analysis}\label{sect:shap}
The advantage of an NN-based accurate predictor is not only the capability of inverting the original pipeline but also the extremely small computational cost and the differentiability of the output features with respect to the input features. For this reason, the NN allows a large number of evaluations to explore how the input (the spectra) affects the output (the parameters). To this aim, we employ the SHapley Additive exPlanations\footnote{\url{https://github.com/shap/shap}, version 0.46.0} (SHAP), which connects optimal credit allocation with local explanations using the classical Shapley values from game theory and their related extensions \citep{Lundberg2017}.
More in detail, this cooperative game theory concept provides a fair distribution of the prediction to the input data depending on their contribution to the global prediction. To calculate the SHAP value for a specific input feature (in our case, a spectral velocity channel), we compute the average marginal contribution of that channel across all possible combinations of channels. This considers the importance of the input channel both alone and in all possible combinations.

We employ the class DeepExplainer of SHAP, based on DeepLIFT \citep{Shrikumar2017}, that  approximates the conditional expectations of SHAP values using a selection of background samples. Note that DeepExplainer effectively provides local feature contributions for specific predictions, but it does not meet the criteria for global sensitivity analysis, which requires evaluating input impact across the entire input space \citep{Saltelli2017}. We apply this method to each parameter independently using the models in the training set to determine the channels' contributions to the prediction of that specific parameter.
The SHAP values are then evaluated on the test set. 

\begin{figure*}
    \centering
    \includegraphics[width=0.98\textwidth]{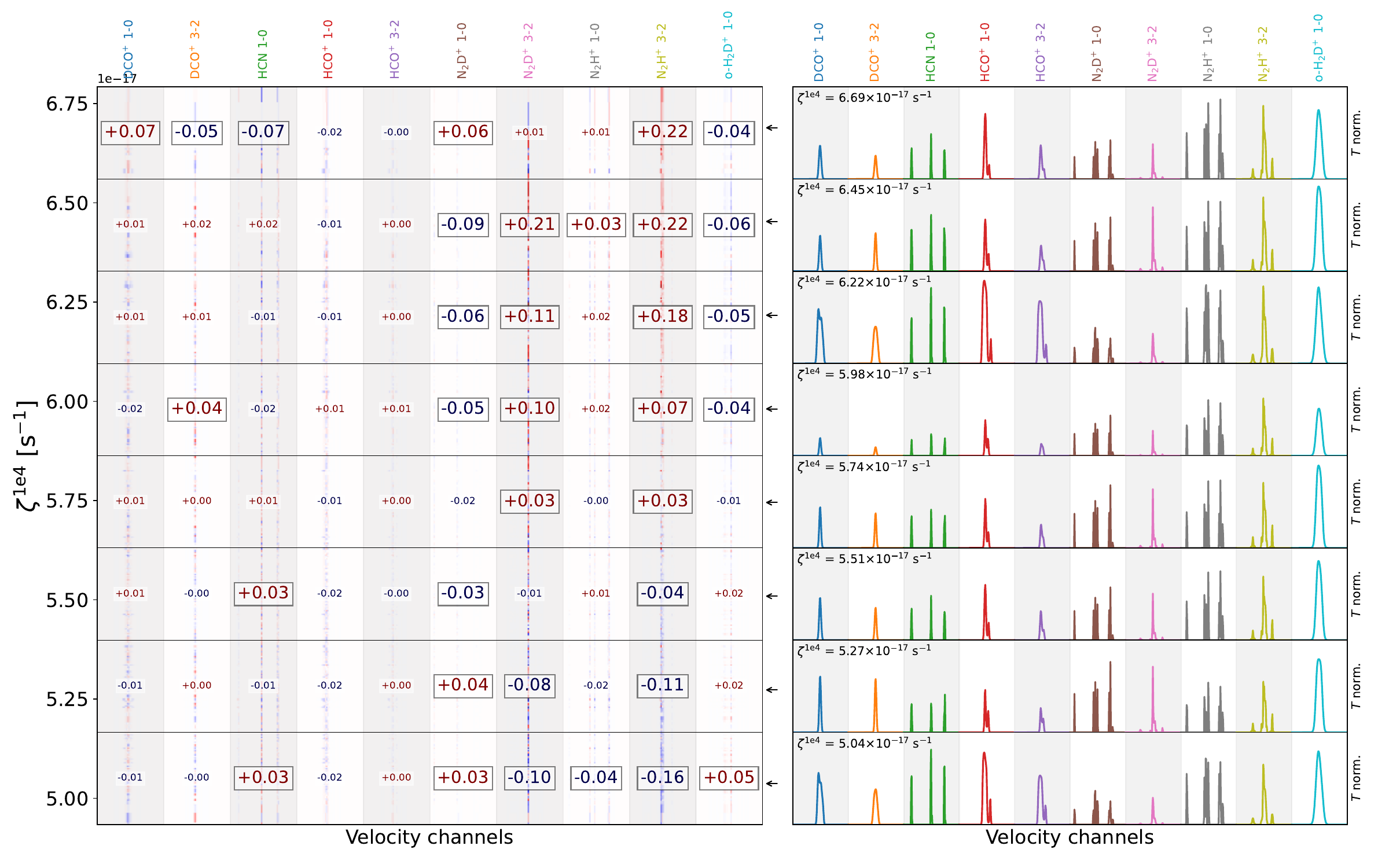}
    \caption{Left panel: the colormap indicates the SHAP values for each channel and each $\zeta^{1e4}$ value in the test set. The numbers indicate the average of the SHAP value in the given part of the plot. Numbers with bounding boxes have an absolute value larger than 0.03. Right panels: Spectra corresponding to the model at the $\zeta^{1e4}$ value indicated by the arrow (and reported in the upper left part of each panel). For the sake of clarity, the temperature is normalized to the maximum of each transition (similarly to the NN input data normalization). Each color corresponds to a different molecular transition, as reported by the labels at the top. The labels also indicate the correspondence between the channels in the left and right panels.} \label{fig:shap-crate}
\end{figure*}

This analysis allows us to produce a figure similar to \fig{fig:shap-crate} for each parameter. The large left panel shows a colormap of the SHAP values of the test set models. On the $x$-axis, we have the velocity channels, while on the $y$-axis, the parameter value of the given test set. To compute the numbers distributed over the left panel, we average each channel's SHAP values and take the maximum. In this way, we have the maximum average contribution of the given transition in the range of parameter values between the two horizontal solid lines. For the sake of clarity, if the absolute value of the maximum is above $0.03$, we render the number with a larger font and a bounding box. Otherwise, the numbers are represented with a smaller font. The font color indicates the positive (red) and negative values (blue). In the right small panels, we have the spectra of some selected models corresponding to the positions indicated by the arrows outside the left panel. The position of the arrow marks the exact value on the $y$-axis and corresponds to the text in the top left corner of each panel on the right.

\begin{figure*}
    \centering
    \includegraphics[width=0.98\textwidth]{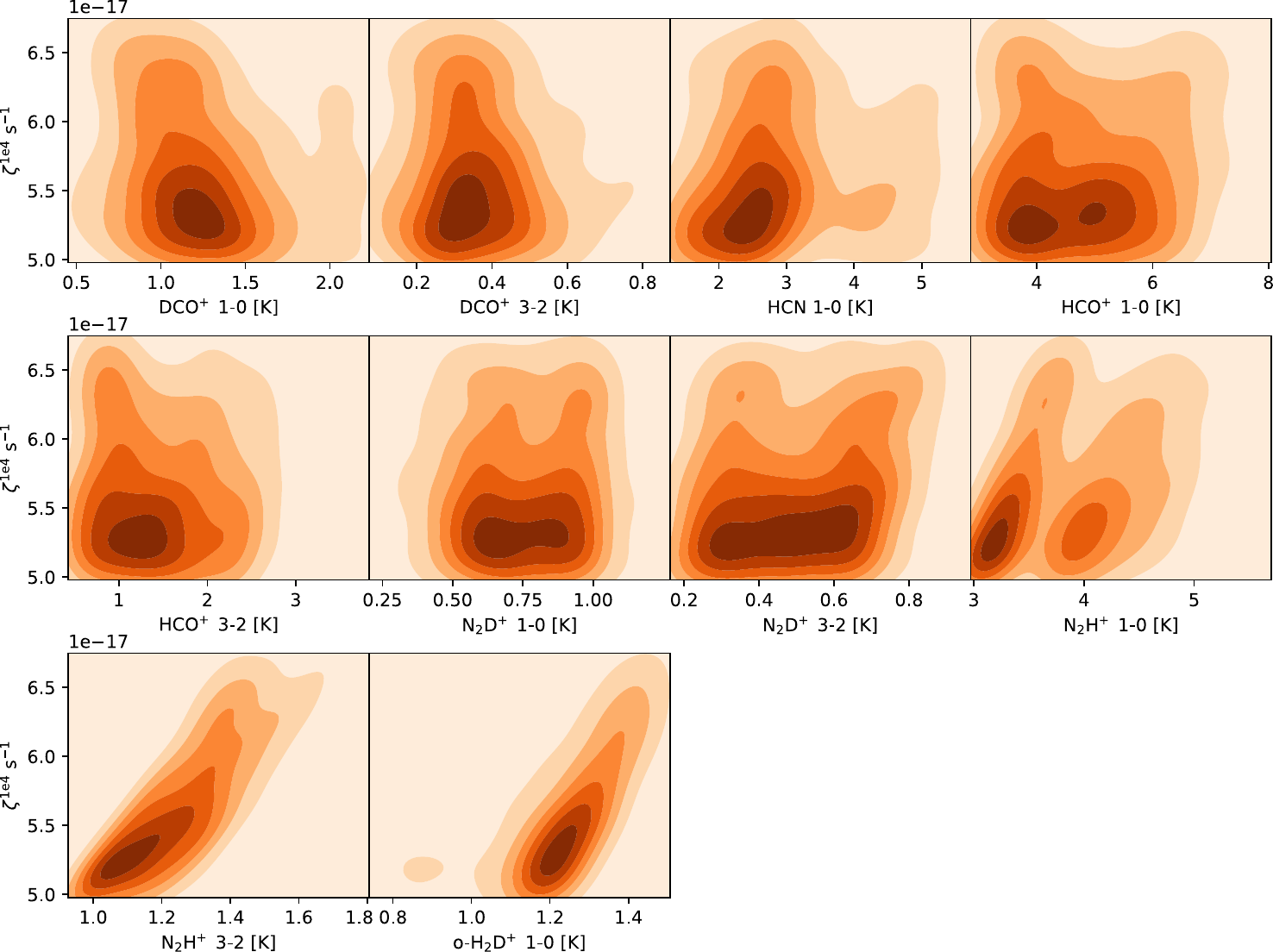}
    \caption{KDE of the peak value of the spectrum for each transition (one per panel, see $x$-axis label) and the  $\zeta^{1e4}$ of the corresponding model (see $y$-axis label). Note the scaling factor for the $y$-axis (1e-17), and the vaying range of the $x$-axis. The parameters of the KDE algorithm are the same as in \fig{fig:kde}.} \label{fig:kde-crate}
\end{figure*}

From this figure, we find that the 3-2 transition of \ce{N2H+} and both transitions of \ce{N2D+} (3-2 more than 1-0) are the key contributors to the value of the cosmic-ray ionization rate at $10^4$\,au ($\zeta^{1e4}$). The sign of the average SHAP value indicates the positive or negative contribution to the parameter, e.g., a large value of the \ce{N2H+} (3-2) transition produces high values of $\zeta^{1e4}$ (positive values), and vice-versa. This behavior corresponds to the right panels, where the peak of \ce{N2H+} (3-2) spectrum scales proportionally to $\zeta^{1e4}$.

The SHAP analysis also indicates which parts of the spectrum contribute to the final value. For example, \ce{N2H+} (3-2) has a broader vertical strip of SHAP values in the left panel, when compared to \ce{N2D+} (3-2); Analogously, HCN contributes with all the three hyperfine lines, \ce{N2D+} (1-0) with two lines, and o-\ce{H2D+} with the ``sides'' of the spectral line, suggesting that the width of the line plays a role in the final result. However, the contributions of these two molecules are negligible, meaning that their presence is only needed to ``fine-tune'' the NN prediction.

Conversely, \ce{HCO+} seems to play a minor role. However, from \citet{Caselli1998}, we know that \ce{HCO+} and \ce{DCO+} are powerful indicators of the cosmic-ray ionization rate. This is because SHAP method is an \emph{explainer} of the interpretation of the NN. The optimization found by the NN is accurate, i.e., it predicts the correct values (see \fig{fig:kde}), but it is not necessarily the most straightforward explanation\footnote{In other words, the NN can be described as a complicated fitting function. For example, a 10-degree polynomial that fits the data of a black body emission is accurate, but it is not the simplest (and physically meaningful) function. Conversely, the fit with a Planck function has one parameter, the temperature, and a physical interpretation.}. This is clear from \fig{fig:kde-crate}, where we plot the KDE of the maximum intensity of each transition against the expected $\zeta^{1e4}$ value in the test set. We note that \ce{N2H+} (3-2) correlates with the ionization rate, as suggested by the SHAP analysis. However, from \fig{fig:kde-crate} (last panel), the peak of o-\ce{H2D+} also seems to correlate with $\zeta^{1e4}$, while SHAP, from \fig{fig:shap-crate} (last column of the first panel), ``suggests'' that most of the information resides in the width of the transition rather than in the peak. Conversely, the role of \ce{N2D+} (3-2) is unclear from the KDE plot alone, while SHAP gives this further insight.

\begin{figure*}
    \centering
    \includegraphics[width=0.98\textwidth]{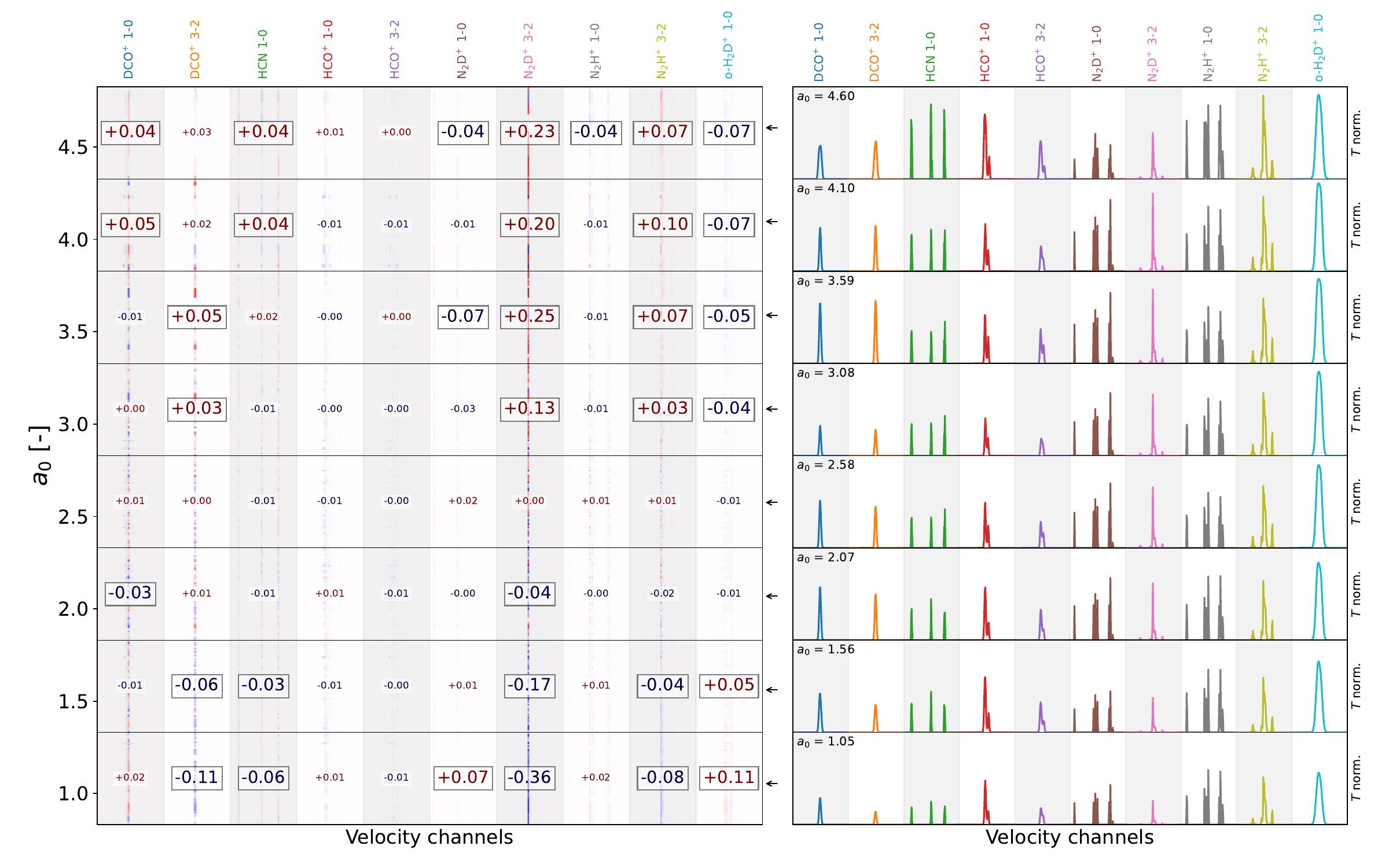}
    \caption{Same as \fig{fig:shap-crate} but for the the $a_0$ parameter of the $\zeta(r)$ fitting function in \eqn{eqn:crfit}.} \label{fig:shap-a0}
\end{figure*}

\begin{figure*}
    \centering
    \includegraphics[width=0.98\textwidth]{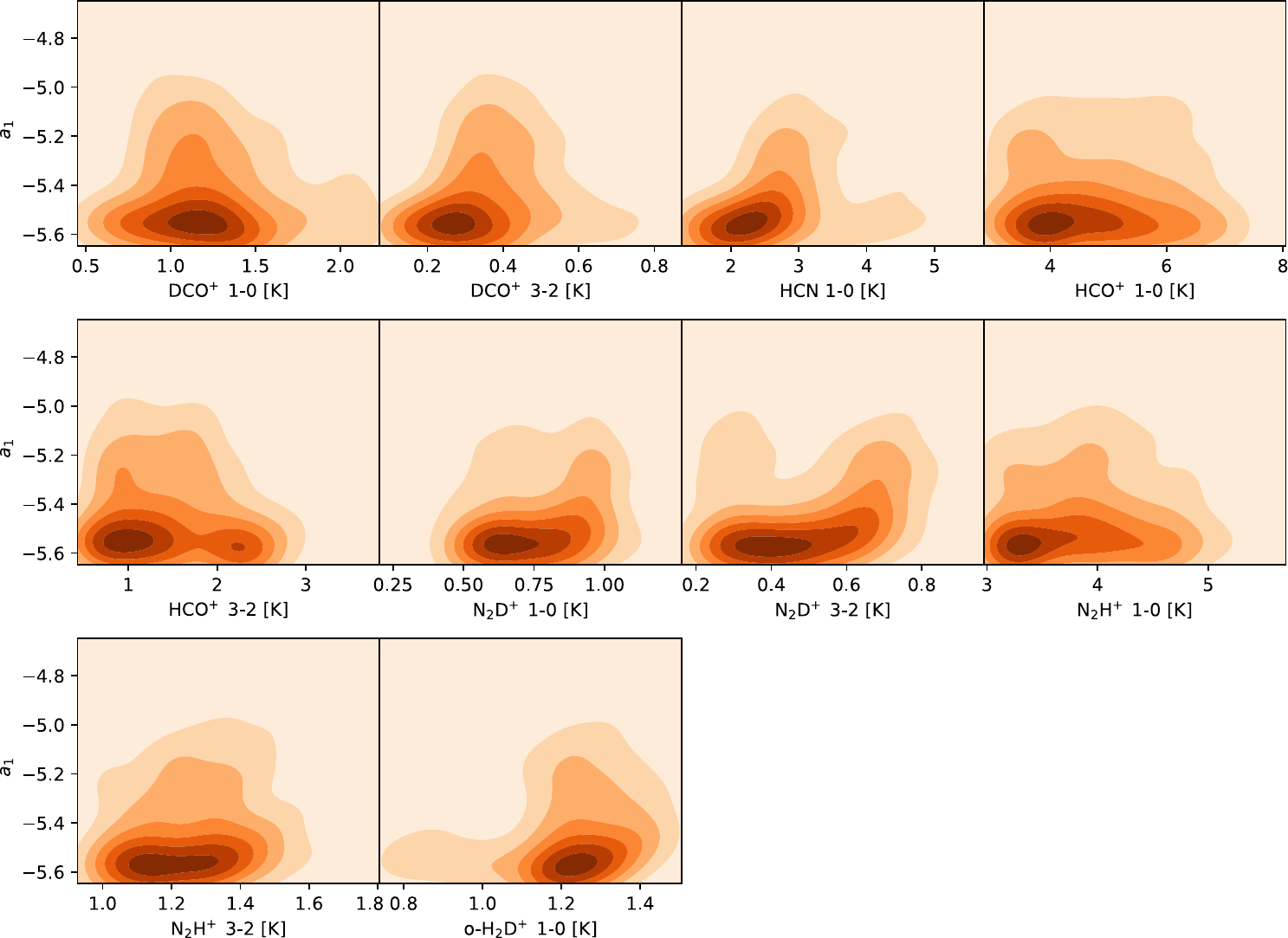}
    \caption{Same as \fig{fig:kde-crate} but for the the $a_1$ parameter of the $\zeta(r)$ fitting function in \eqn{eqn:crfit}.} \label{fig:kde-a1}
\end{figure*}

The NN allows us to determine single quantities, such as $\zeta^{1e4}$, but also multiple components at once from the observations, like predicting the four coefficients $a_i$ of the fit we described in \eqn{eqn:crfit}. Here, the interpretation is more complicated since several transitions play a different role for different coefficients and different coefficient values. As an explanatory case, we discuss $a_0$, which represents the scaling of the fitting function, and hence, it is immediately connected to the average value of cosmic-ray ionization rate. We note from \fig{fig:shap-a0} that $a_0$ is affected by \ce{N2D+} (3-2) in a similar way to  $\zeta^{1e4}$. However, other molecules (\ce{DCO+}, HCN, \ce{N2H+}, and o-\ce{H2D+}) are also involved in determining $a_0$, especially closer to its upper and lower values. Analogously, the other coefficients present similar complex interplay patterns, especially $a_1$, which also has the largest uncertainty, being the scaling factor in the argument of an exponential function. The importance of this machine learning-based analysis is also suggested by \fig{fig:kde-a1}, where $a_1$ does not show any obvious correlation with the peaks of the lines.

\begin{figure*}
    \centering
    \includegraphics[width=0.98\textwidth]{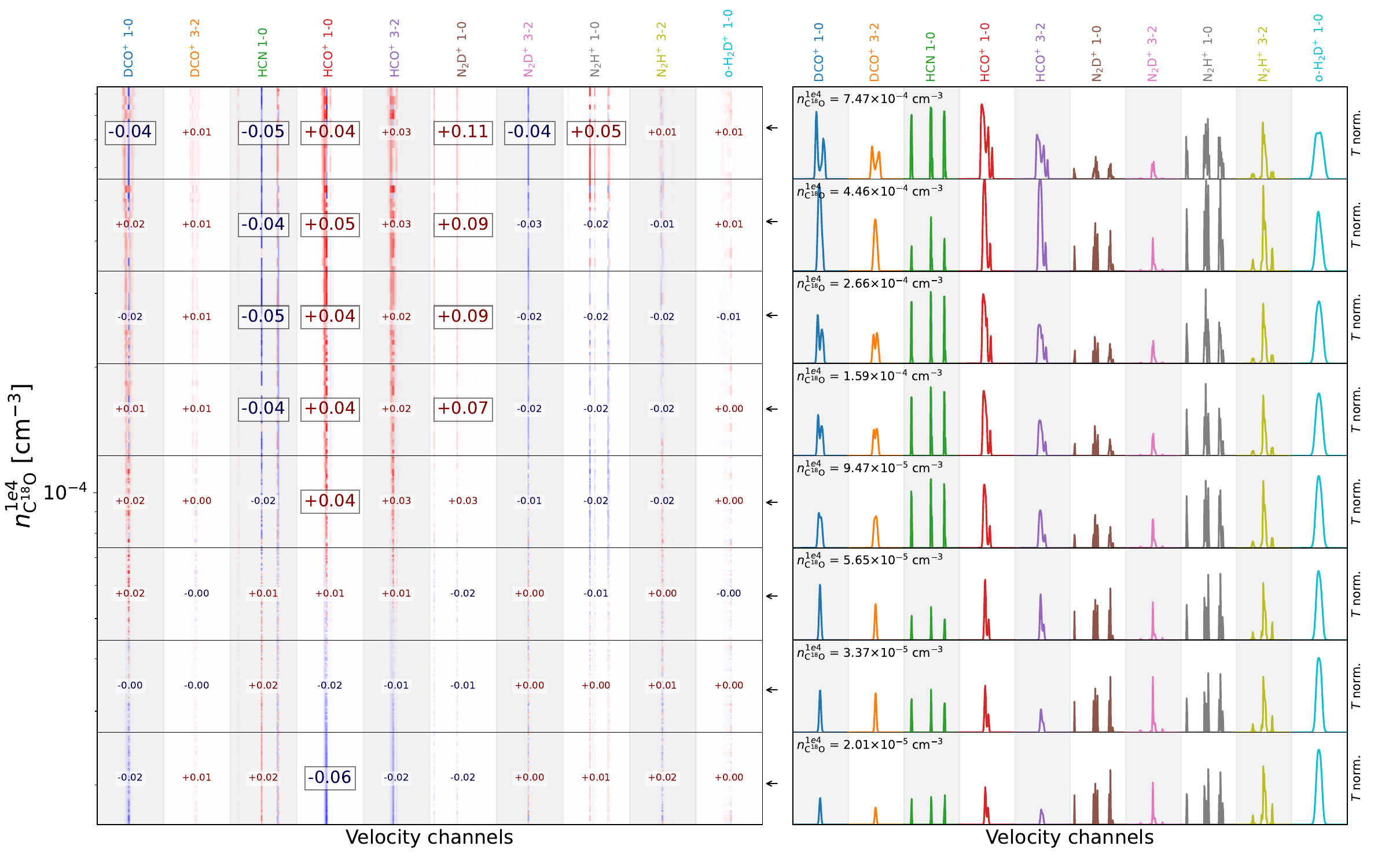}
    \caption{Same as \fig{fig:shap-crate} but for the abundance of \ce{C^18O} at $10^4$\,au ($n_{\ce{C18O}}^{1e4}$).} \label{fig:shap-C18O}
\end{figure*}

\begin{figure*}
    \centering
    \includegraphics[width=0.98\textwidth]{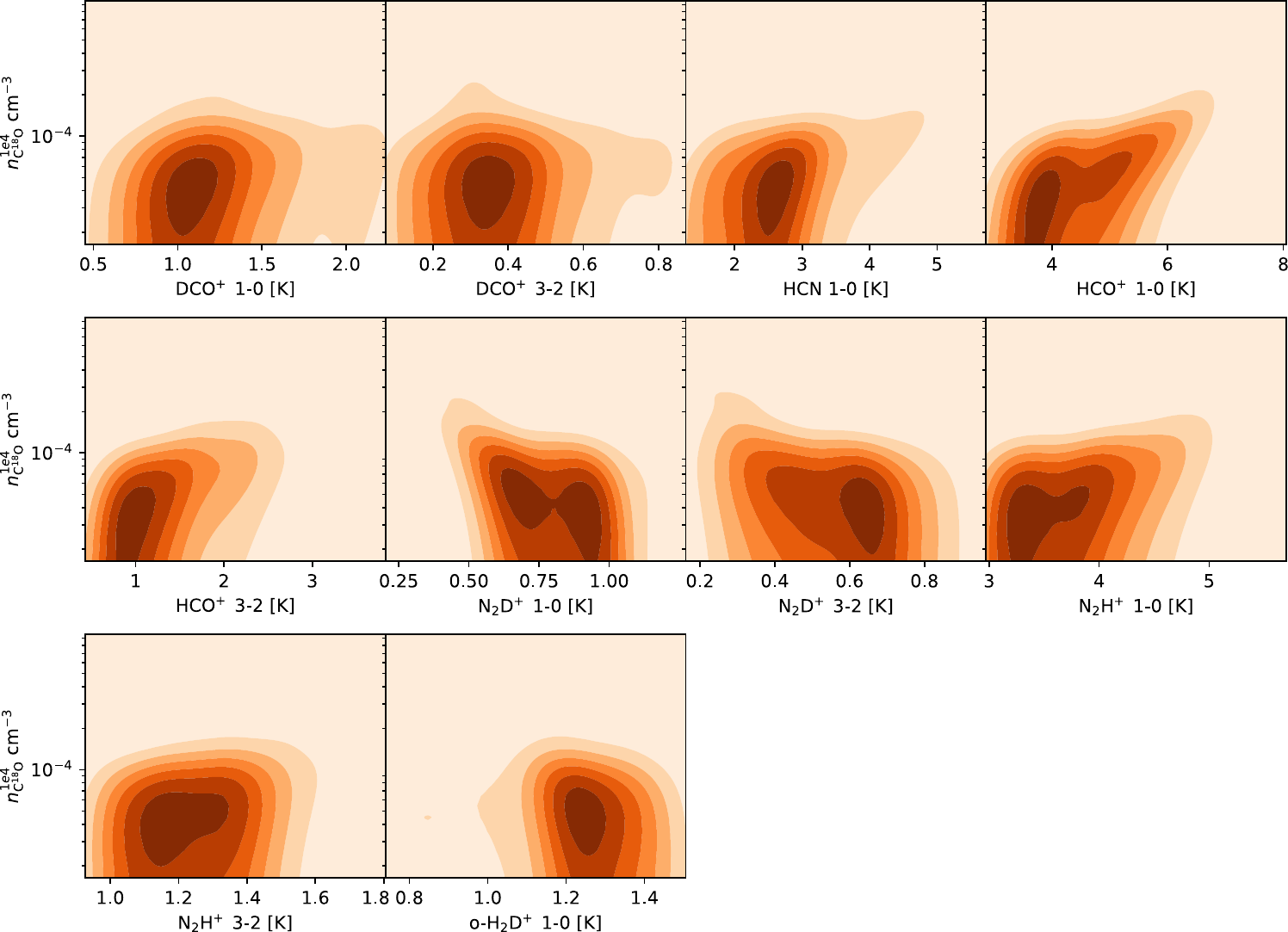}
    \caption{Same as \fig{fig:kde-crate} but for the abundance of \ce{C^18O} at $10^4$\,au ($n_{\ce{C18O}}^{1e4}$).} \label{fig:kde-C18O}
\end{figure*}

In addition, using the NN could give information on quantities that are not derivable from the observation alone. For example, \ce{C^18O} at $10^4$\,au ($n_{\ce{C^18O}}^{1e4}$) is well-reproduced by the NN, although we do not include any \ce{C^18O} emission line in the input of the NN, thus suggesting that the other lines can fairly predict it. In this case, the SHAP analysis (\fig{fig:shap-C18O}) indicates that \ce{N2D+} (1-0), \ce{HCO+} (1-0), and HCN retain the information necessary for the NN to reconstruct the abundance of \ce{C^18O}. If we compare this result with the KDE distribution of the peaks of the lines (\fig{fig:kde-C18O}), we note that in this case, there is no clear trend to determine \ce{C^18O}, indicating that is the combined information from all the lines that is used to predict the result. The prediction of \ce{C^18O} is possible because these species are interconnected, both directly (e.g., \ce{HCO+} is a key component of the CO formation network) and indirectly (i.e., similar physical parameters have the same impact on different molecules). However, we note that in our chemical network, the abundance of \ce{C^18O} is scaled from that of CO, suggesting that actual information is related to CO rather than \ce{C^18O}.

\begin{figure*}
    \centering
    \includegraphics[width=0.98\textwidth]{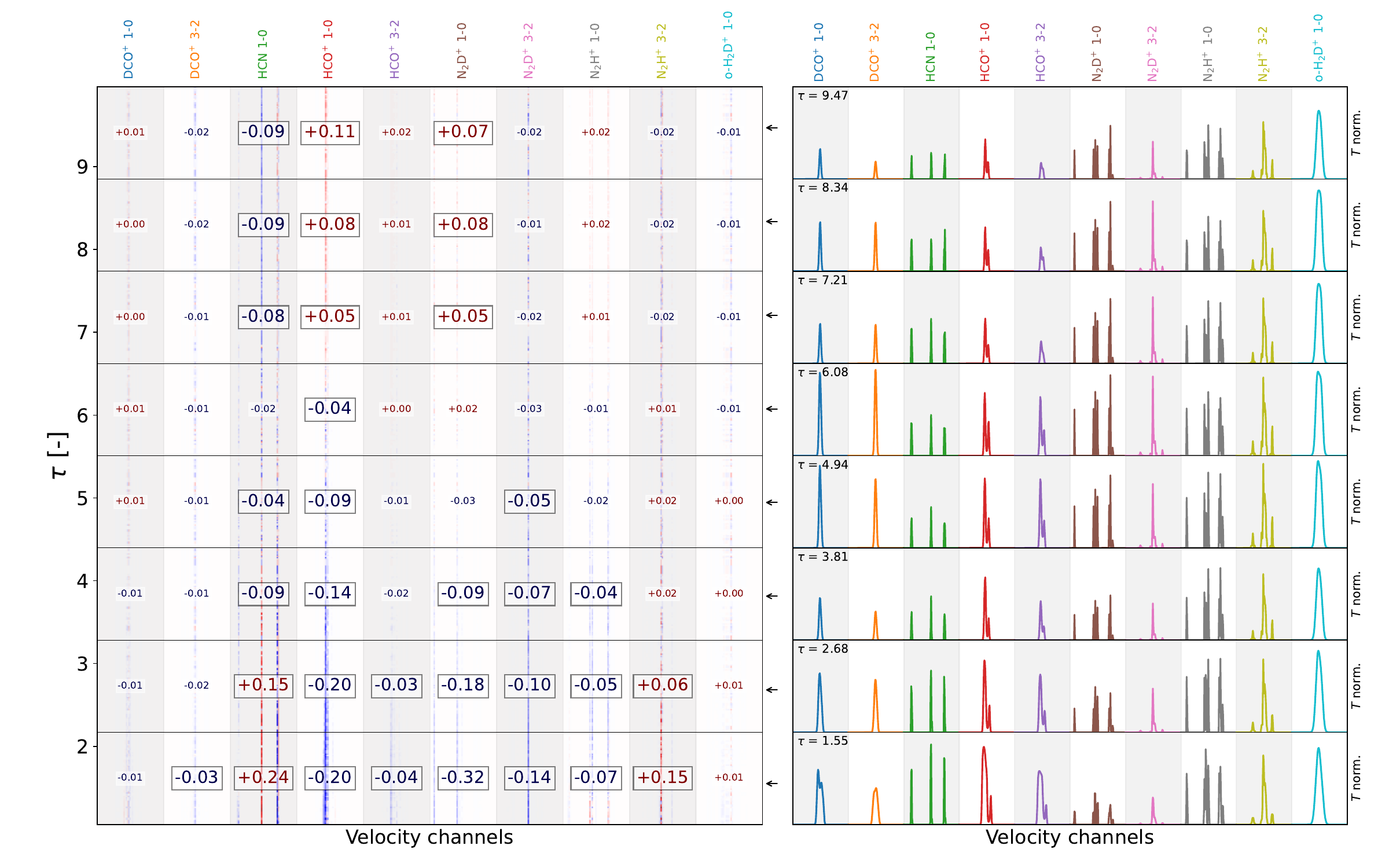}
    \caption{Same as \fig{fig:shap-crate} but for the collapse time factor $\tau$.} \label{fig:shap-tau}
\end{figure*}

We finally discuss the collapse time factor ($\tau$), which is strictly linked to the chemical timescale of the slow-evolving species, for example, HCN (\fig{fig:shap-tau}). The abundance of this species depends on the initial conditions (i.e., N and \ce{N2}) and on the temporal evolution. Since, in our case, the initial conditions are fixed, if we allow a slower collapse (larger $\tau$), HCN is destroyed more rapidly, as indicated by the weakening of the lines in the right panels. A similar behavior is shown by \ce{N2D+}, which increases its abundance when longer times favor the conversion from \ce{N2H+}.

In principle, it is possible to perform this analysis on every parameter and employ additional parameters not included at this stage (e.g., specify quantities at different radii than $10^4$\,au, or define fitting functions similar to Eq.\,\ref{eqn:crfit}). However, for the aims of the present work, we limit our analysis to the aforementioned quantities.

\section{Limitations}\label{sect:limitations}
Models are limited by definition. The choice of the 1D profile not only limits the modeling of any geometrical asymmetry but does not allow the inclusion of any magnetic field. In this regard, the choice of parameterizing the presence of magnetic fields with $\tau$ does not completely solve the problem and limits the interpretability of the collapse dynamics. Even with additional physics, it is important to remark that the Larson collapse is a simplified approach with respect to the observed dynamics \citep{Keto2015,Vaytet2017}.

Computing the hydrodynamics in the isothermal case first and then computing the temperature on top of the tracer particles is another assumption that might exclude any complex interplay between hydrodynamics and thermochemical evolution. However, the temperature range we found suggests that \emph{within this context} this interplay is limited. Despite the use of tracer particles, which removes the issue of chemical species advection, post-processing might produce inaccurate results due to the lack of self-consistency \citep{FerradaChamorro2021}. However, there are successful cases where it is consistently employed (e.g., \citealt{Panessa2023}).

Although the ``large'' chemical network contains 136 chemical species, it does not include several species that might play a role in the synthetic observations or in the chemical evolution, such as ammonia or \ce{C^18O} \citep{Sipila2022}. In addition, we limit our analysis to fixed initial chemical conditions, but some slowing-evolving species like HCN depend on the specific initial conditions \citep{HilyBlant2010}.
Similarly, some of the correlations found or some of the observed features might be produced by non-chemical correlations generated by some inaccurate chemical rate, missing chemical pathway, or time-dependent effect enhanced by the inaccurate initial conditions. In general, synthetic observations observe the modeled chemical network rather than an actual astrophysical object. Therefore, they should be interpreted carefully, especially when large and complex chemical networks are involved, and multiple species are observed simultaneously.

\tg{}{A comprehensive review of the techniques to link observations and physical quantities is beyond the aims of the present paper. Therefore, we limit our summary to some of the standard and well-established implementations that include best model parameter grid search based on metric minimization of observed and simulated quantities (e.g., \citealt{Sheffer2011}, \citealt{Joblin2018}, and \citealt{Wu2018}), and methods that do not rely on pre-computed grids, such as gradient descent with various algorithms (e.g., \citealt{Galliano2003}, \citealt{Marchal2019}, and \citealt{Paumard2022}), Bayesian-based model selection (e.g., \citealt{Zucker2021}, \citealt{Chevallard2016}, \citealt{Johannesson2016}, and \citealt{Behrens2022}), and Monte Carlo sampling with a large variety of implementations (e.g., \citealt{Chevallard2013}, \citealt{Makrymallis2014}, \citealt{Gratier2016}, \citealt{Galliano2018}, \citealt{Holdship2018}, \citealt{Galliano2021}, and \citealt{Ramambason2022}).

We will not review the advantages and limitations of each technique here due to the problem's specific details, the astrophysical object(s) of interest, the physical model developments, and the technical implementations of the various methodologies. We merely point out that some of these techniques are capable of producing estimates of uncertainty, unlike our method (at least at the present stage). Furthermore, a common disadvantage is the \emph{intrinsic} parameter degeneracy that limits all the aforementioned techniques, including the one presented in this paper. In our case, the forward operator (i.e., stages 1-6 in \fig{fig:scheme}) is not necessarily invertible, so the parameters' degeneracy might limit the optimization of the NN. In addition, if newer transitions are available or some are missing, the NN requires additional training that might not necessarily converge. On the other hand, in addition to the differentiability and the seamless interface with the SHAP architecture, the use of an NN presents some advantages, such as a training stage of a couple of minutes on a standard GPU and a negligible evaluation time, making this technique extremely effective and flexible.}

\tg{}{To further discuss the limitations, we present an application to the prestellar core L1544. However, we remind the reader that some technical aspects currently limit the applicability, and they can be addressed in the future.} As we discussed earlier, emulators and NN are limited to a specific feature domain, and their application to real observations might produce false correlations or emulator hallucinations. In principle, it is possible to use the backward model $\mathcal{F}^{-1}$ to determine the physical characteristics of an observed object. \tg{For example, for}{In} the case of the prestellar core L1544, the transitions observed cover several molecules available in our model, in particular,  \ce{DCO+} (1-0, 3-2, \citealt{Redaelli2019a}), HCN (1-0, \citealt{HilyBlant2010}), \ce{HCO+} (1-0, 3-2, \citealt{Redaelli2022b}), \ce{N2D+} (1-0, 3-2, \citealt{Redaelli2019a}), \ce{N2H+} (1-0, 3-2, \citealt{Redaelli2019a}), and o-\ce{H2D+} (1-0, \citealt{Caselli2003,Vastel2006}). We interpolate the observed line profiles to the channels we used in the training set, assuming zero intensity outside. However, some discrepancies exist between the simulated model and the observed quantities. The most relevant is underestimating of \ce{C^18O} and \ce{DCO+} line intensities. In addition, the models with the largest intensities of \ce{C^18O} show a double peak, which is absent in the case of L1544  (see also \citealt{Keto2015} showing that quasi-equilibrium contraction is needed to reproduce the \ce{C^18O} and \ce{H2O} lines in L1544, while Larson-Penston and singular isothermal sphere models fail). \tg{}{Additionally, our implementation might fail on L1544 because the NN is not trained with a consistent noise model. However, we need more tests to verify whether this issue is relevant in this context.}

With these premises, the reason why we do not obtain reliable results at this stage, is mainly the presence of unexpected features in the input spectra. This artificially enhances or decreases some input nodes that generate anomalous output. The results are also influenced by the fact that probably none of the generated models resemble L1544, or that, given the complexity of our pipeline, any of the ingredients might determine an NN hallucination. However, despite the limited results, the approach of using a backward emulator is still promising, also considering different NN architectures that might be more appropriate for spectral data, like convolutional neural networks \citep{Kiranyaz2019}, and transformers \citep{Leung2024,Zhang2024}. Another approach could be to provide the NN with the actual fitting parameters of the lines using standard methods instead of the raw spectra (e.g., \citealt{Ginsburg2022}).

\section{Conclusions}\label{sect:conclusions}
In this work, we proposed a pipeline to connect the information in 1D hydrodynamical models of prestellar cores with the observed features of their synthetic spectra. This procedure employs 3000 models of hydrodynamical gravitational collapse with time-dependent thermochemistry, isotopologue chemistry, and consistent cosmic-ray propagation. We use SHAP, an interpretable machine learning technique, to connect the models' parameters and the corresponding generated synthetic spectra.

The main findings are:
\begin{itemize}
    \item Most of the information is retained by the spectra, apart from the small turbulence velocity dispersion ($\sigma_\varv$) and the initial total mass ($M$).
    \item SHAP, when applied to synthetic observations, represents a valid tool to explore how spectral features are connected to model parameters.
    \item Within the assumptions of our model, cosmic rays are well reproduced. Their ionization rate at a distance of $10^4$\,au is retrieved mainly by \ce{N2H+} and \ce{N2D+}, while the radial profile $\zeta(r)$ is determined by a mix of the contributions of \ce{N2D+}, \ce{DCO+}, HCN, \ce{N2H+}, and o-\ce{H2D+}.
    \item This method is capable of obtaining information on the chemistry of molecules not included in the spectra. To this aim, we have removed \ce{C^18O} from the spectra. We found that a combination of \ce{N2D+}, \ce{HCO+}, and HCN can be used to constrain the abundance of \ce{C^18O} at $10^4$\,au.
    \item The effects of the time dependence, namely the collapse slowing factor $\tau$, emerge in the features of the HCN spectra, having a relatively slow chemical time scale.
    \item Future work is required to address the role of model uncertainties, neural network limitations, and backward emulation applied to real observations.
\end{itemize}

\begin{acknowledgements}
\tg{}{We thank the referee, Pierre Palud, for improving the quality of the paper.}
 We thank Mika Juvela for the LOC update to include radially-dependent collider abundances. We thank Johannes Heyl for suggesting using DeepExplainer instead of KernelExplainer.  This research was supported by the Excellence Cluster ORIGINS, funded by the Deutsche Forschungsgemeinschaft (DFG, German Research Foundation) under Germany's Excellence Strategy -- EXC-2094 -- 390783311.
 DG acknowledges support from the INAF minigrant PACIFISM. SB acknowledges BASAL Centro de Astrofisica y Tecnologias Afines (CATA), project number AFB-17002.
 SB acknowledges BASAL Centro de Astrofisica y Tecnologias Afines (CATA), project number AFB-17002. This research has made use of spectroscopic and collisional data from the EMAA database (\url{https://emaa.osug.fr} and \url{https://dx.doi.org/10.17178/EMAA}).
\end{acknowledgements}

\bibliographystyle{aa}
\bibliography{mybib}

\begin{appendix}
\section{Chemical species included in the ``small'' chemical network}\label{sect:small-network}
 The ``small'' chemical network includes the following chemical species: \ce{e-}, \ce{H}, \ce{H+} \ce{H-}, \ce{He}, \ce{He+}, \ce{He++}, \ce{C}, \ce{C+}, \ce{C-}, \ce{O}, \ce{O+}, \ce{O-}, \ce{H2}, \ce{H2+}, \ce{H3+}, \ce{OH}, \ce{OH+}, \ce{CH}, \ce{CH+}, \ce{CH2}, \ce{CH2+}, \ce{CH3+}, \ce{H2O}, \ce{H2O+}, \ce{H2O}$_{\rm d}$ \ce{HCO}, \ce{HCO+}, \ce{HOC+}, \ce{CO}, \ce{CO+}, \ce{CO}$_{\rm d}$, \ce{O2}, \ce{O2+}, \ce{C2}, and \ce{H3O+}, where the subscript ``d'' indicates the species on ice. We employ the rate coefficients taken from \citet{DeJong1972,Aldrovandi1973,Poulaert1978,Karpas1979,Mitchell1983,Lepp1983,Janev1987,Ferland1992,Verner1996,Abel1997,Savin2004,Yoshida2006,Harada2010,Kreckel2010,Grassi2011a,Forrey2013,Stenrup2009,OConnor2015,Wakelam2015,Millar2024}.   More details in \citet{Glover2010} and in the \textsc{KROME} network file \texttt{react\_CO\_thin\_ice}, commit \texttt{6a762de}.

\section{Chemical species included in the ``large'' chemical network}\label{sect:large-network}
The ``large'' chemical network includes: \ce{e-}, \ce{H}, \ce{H+}, \ce{H-}, \ce{D}, \ce{D+}, \ce{D-}, o-\ce{H2}, p-\ce{H2}, o-\ce{H2+}, p-\ce{H2+}, \ce{HD}, \ce{HD+}, o-\ce{H3+}, p-\ce{H3+}, \ce{He}, \ce{He+}, o-\ce{D2}, p-\ce{D2}, o-\ce{D2+}, p-\ce{D2+}, o-\ce{H2D+}, p-\ce{H2D+}, \ce{HeH+}, o-\ce{D2H+}, p-\ce{D2H+}, m-\ce{D3+}, o-\ce{D3+}, p-\ce{D3+}, \ce{C}, \ce{C+}, \ce{C-}, \ce{CH}, \ce{CH+}, \ce{N}$_{\rm d}$, \ce{N}, \ce{N+}, \ce{CD}, \ce{CD+}, \ce{CH2}, \ce{CH2+}, \ce{NH}, \ce{NH+}, \ce{CHD}, \ce{CHD+}, \ce{O}$_{\rm d}$, \ce{O}, \ce{O+}, \ce{O-}, \ce{ND}, \ce{ND+}, o-\ce{NH2}, p-\ce{NH2}, o-\ce{NH2+}, p-\ce{NH2+}, \ce{CD2}, \ce{CD2+}, \ce{OH}, \ce{OH+}, \ce{OH-}, \ce{NHD}, \ce{NHD+}, \ce{OD}, \ce{OD+}, \ce{OD-}, o-\ce{H2O}, p-\ce{H2O}, o-\ce{H2O+}, p-\ce{H2O+}, o-\ce{ND2}, p-\ce{ND2}, o-\ce{ND2+}, p-\ce{ND2+}, \ce{HDO}, \ce{HDO+}, o-\ce{D2O}, p-\ce{D2O}, o-\ce{D2O+}, p-\ce{D2O+}, \ce{C2}, \ce{C2+}, \ce{C2H+}, \ce{CN}, \ce{CN+}, \ce{CN-}, \ce{C2D+}, \ce{HCN}, \ce{HNC}, \ce{HCN+}, \ce{HNC+}, \ce{CO}$_{\rm d}$, \ce{CO}, \ce{CO+}, \ce{N2}$_{,\rm d}$, \ce{N2}, \ce{N2+}, \ce{DNC}, \ce{DCN}, \ce{DNC+}, \ce{DCN+}, \ce{HCO}, \ce{HCO+}, \ce{HOC+}, \ce{N2H+}, \ce{NO}, \ce{NO+}, \ce{DCO}, \ce{DOC+}, \ce{DCO+}, \ce{N2D+}, \ce{HNO}, \ce{HNO+}, \ce{O2}, \ce{O2+}, \ce{DNO}, \ce{DNO+}, \ce{O2H}, \ce{HO2+}, \ce{O2D}, \ce{DO2+}, \ce{C3}, \ce{C3+}, \ce{C2N+}, \ce{CNC+}, \ce{CCO}, \ce{C2O+}, \ce{OCN}, \ce{NCO+}, \ce{CO2}, \ce{CO2+}, \ce{N2O}, \ce{NO2}, \ce{NO2+}, GRAIN, GRAIN$^+$, and GRAIN$^-$, where o-, p-, and m-, indicate the ortho-, para-, and meta- spin states, respectively. For further details on the network, we refer the reader to \citet{Bovino2020}.

\section{Model comparison}\label{sect:comparison}
\begin{figure*}
    \centering
    \includegraphics[width=0.98\textwidth]{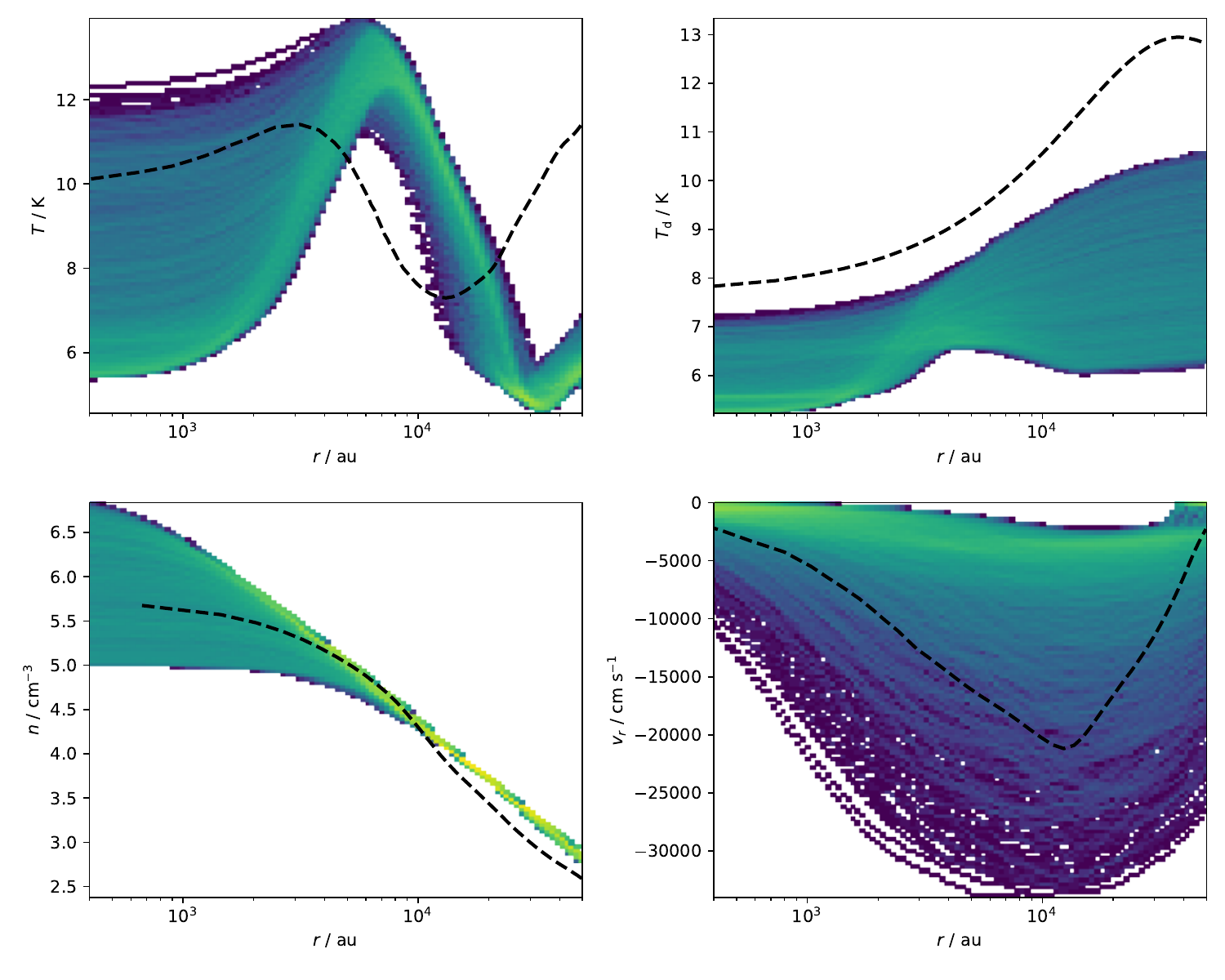}
    \caption{Radial profiles for different quantities (gas temperature $T$, dust temperature $T_{\rm d}$, density $n$, and $\varv_r$ radial velocity) as density probability of our models' sample (color plot), compared with Fig.\,3 of \citet{Sipila2022}, at $t=7.19\times10^5$\,yr curves (dashed lines). The probability density is vertically normalized to have the same integral value (i.e., unity) at each radius. Note the log-scaled radius.} \label{fig:comparison}
\end{figure*}

In \fig{fig:comparison}, we compare the radial profiles of various quantities in our models to \citet{Sipila2018} (gas temperature $T$, dust temperature $T_{\rm d}$, density $n$, and $\varv_r$ radial velocity). The main differences in gas temperature ($T$) are due to different assumptions on the heating of cosmic rays (where we use \citealt{Glassgold2012} implementation and variable cosmic-ray ionization rate), and CO cooling (where we employ a custom look-up table from \citealt{Omukai2010}), while \citet{Sipila2018} use \citet{Juvela1997}. In addition, the size of the computational domain plays a role in determining the CO column density. The difference in dust temperature ($T_{\rm d}$) is produced by a different assumption on the grain optical properties and again on the limit of the computational domain. The radial velocity ($\varv_r$) is statistically smaller due to the arbitrary scaling of the collapse time factor ($\tau$). These differences do not significantly influence the main findings and the method description.

\section{Forward emulation: from models to observations} \label{sect:forward}

Analogously to the backward emulation, we trained an NN that predicts the observed spectra from the model parameters, i.e., a $\mathbf{s}=\mathcal{F}(\mathbf{p})$ emulator. This allows us to generate spectra extremely fast by simply specifying arbitrary parameters. 

\tg{}{In this case, the NN architecture is a fully connected feedforward consisting of two hidden layers (64 and 32 neurons). Each hidden layer has a ReLU activation function. The NN is optimized via an Adam optimizer with a learning rate of $10^{-3}$, and the mean squared error (MSE) loss function is utilized to compute the loss. The other options are the default \textsc{Pytorch} options. We trained the network for 6000 epochs.}

With respect to the previous method, the training efficiency is now very high, and all the line features are well-reproduced, even in the test set. This behavior is due to the lack of large degeneracies in the forward model.

After the training, it is possible to use the NN to predict any combination of values, including the non-physical ones, i.e., exploring the white (no data) parts in the histograms of the lower left triangle of the matrix of plots in \fig{fig:correlation}. For example, the NN can produce all the spectra for a hypothetical model with a small value of the visual extinction at the limits of the computational domain ($A_{\rm V,0}$) and a small dust temperature at $10^4$\,au ($T_{\rm d}^{1e4}$). Our model does not achieve this combination for physical reasons (i.e., high $A_{\rm V,0}$ means insufficient radiation to heat the dust grains, hence a smaller $T_{\rm d}^{1e4}$).

This approach is usually not recommended even if, in some cases, this ``extrapolative'' technique could lead to alluring physical interpretations. In fact, as a cautionary tale, reducing $\tau$ below the limits first produces a physical reduction of the HCN emission (being the HCN abundance relatively time-dependent), but then generates unphysical HCN absorption lines (NN hallucinations), which are never present in the training set (nor in the validation or test sets).
For this reason, we limit our discussion to the backward ``observation to model'' NN.

\section{Cosmic ray fit prediction error}\label{sect:cr-fit}
\begin{figure}
    \centering
    \includegraphics[width=0.48\textwidth]{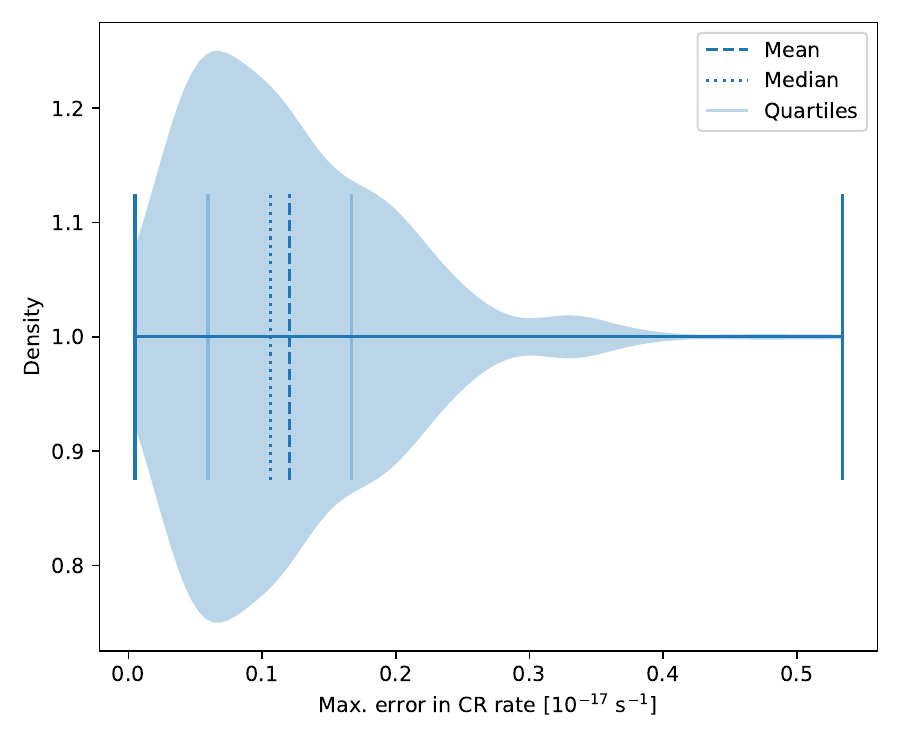}
    \caption{Density probability of the maximum error of predicting $\zeta(r)$ with NN. The mean, the median, and the 25 and 75 quartiles are respectively indicated with dashed, dotted, and solid lines.} \label{fig:cr-fit-errors}
\end{figure}

To quantify the error of the NN in predicting $a_0$, $a_1$, $a_2$, and $a_3$, for each test model, we compute $\max|\hat\zeta(r)-\hat\zeta'(r)|$, where the first term is the actual \eqn{eqn:crfit} and the second is the same equation but using the predicted $a_i$ coefficients. The error probability density distribution is reported in \fig{fig:cr-fit-errors}. The mean, the median, and the 25 and 75 quartiles are respectively indicated with dashed, dotted, and solid lines. The plot shows that the error is well below $5\times10^{-18}$\,s$^{-1}$.

\end{appendix}

\end{document}